\documentclass[
 reprint,
superscriptaddress,
nofootinbib,
 amsmath,amssymb,
pra,
]{revtex4-2}

\usepackage{amsmath}
\usepackage{amssymb}
\usepackage{hyperref,cleveref}
\hypersetup{
    colorlinks=true,
    urlcolor=blue,
    citecolor=blue,
    linkcolor=blue
}
\usepackage{graphicx,float,color}
\usepackage[caption=false]{subfig}

\usepackage{algorithm}
\usepackage[noend]{algpseudocode}
\newcommand{\bC}{\mathbb{C}}
\newcommand{\bI}{\mathbb{I}}
\newcommand{\bZ}{\mathbb{Z}}

\newcommand{\cB}{\mathcal{B}}
\newcommand{\cC}{\mathcal{C}}
\newcommand{\cD}{\mathcal{D}}
\newcommand{\cE}{\mathcal{E}}

\newcommand{\cL}{\mathcal{L}}
\newcommand{\cN}{\mathcal{N}}
\newcommand{\cO}{\mathcal{O}}
\newcommand{\cP}{\mathcal{P}}
\newcommand{\cS}{\mathcal{S}}
\newcommand{\cT}{\mathcal{T}}
\newcommand{\cQ}{\mathcal{Q}}
\newcommand{\cU}{\mathcal{U}}

\renewcommand{\ol}[1]{\overline{#1}}
\newcommand{\Prob}{\text{Prob}}
\newcommand{\tr}{\text{Tr}}
\newcommand{\argmax}[1]{\underset{#1}{\text{argmax}}}

\begin{document}
\title{Enhancing Decoding Performance using Efficient Error Learning}

\author{Pavithran Iyer}
\affiliation{%
Institute for Quantum Computing, University of Waterloo, Waterloo, Ontario N2L 3G1, Canada.
}

\author{Aditya Jain}
\affiliation{
University of Cambridge, The Old Schools, Trinity Ln, Cambridge CB2 1TN, United Kingdom.
}

\author{Stephen D. Bartlett}
\affiliation{%
Centre for Engineered Quantum Systems, School of Physics, University of Sydney, Sydney, New South Wales 2006, Australia.
}

\author{Joseph Emerson}
\affiliation{%
Institute for Quantum Computing, University of Waterloo, Waterloo, Ontario N2L 3G1, Canada.
}
\affiliation{
Department of Applied Mathematics, University of Waterloo, Waterloo, Ontario N2L 3G1, Canada.
}
\affiliation{
Keysight Technologies Canada, Kanata, ON K2K 2W5, Canada.
}

\begin{abstract}
Lowering the resource overhead needed to achieve fault-tolerant quantum computation is crucial to building scalable quantum computers. We show that adapting conventional maximum likelihood (ML) decoders to a small subset of efficiently learnable physical error characteristics can significantly improve the logical performance of a quantum error-correcting code. Specifically, we leverage error information obtained from efficient characterization methods based on Cycle Error Reconstruction (CER), which yields Pauli error rates on the $n$ qubits of an error-correcting code.  Although the total number of Pauli error rates needed to describe a general noise process is exponentially large in $n$, we show that only a few of the largest few Pauli error rates are needed and that a heuristic technique can complete the Pauli error distribution for ML decoding from this restricted dataset. Using these techniques, we demonstrate significant performance improvements for decoding quantum codes under a variety of physically relevant error models. For instance, with CER data that constitute merely $1\%$ of the Pauli error rates in the system, we achieve a $10X$ gain in performance compared to the case where decoding is based solely on the fidelity of the underlying noise process. Our conclusions underscore the promise of recent error characterization methods for improving quantum error correction and lowering overheads.
\end{abstract}

\maketitle

\section{Introduction}
Large-scale quantum computers must be designed to ensure reliable computation in the presence of noise. The theory of fault-tolerant quantum computation makes it possible to encode quantum information and perform an arbitrarily large quantum circuit \cite{S96,P98}, at the expense of a substantial overhead in terms of the number of physical qubits needed to store a logical qubit and the number of physical devices required to execute a logical operation. The overhead poses a significant challenge to realizing the guarantees of fault-tolerant (FT) quantum computation using current hardware. Parallel advances have been made in various aspects of a fault-tolerant scheme to reduce the overhead required \cite{Paetznick14,Gottesman13} to achieve a suppression of noise that affects the logical information, i.e., the logical error rate. These include the development of new quantum error correcting codes \cite{SJCT23,BCGM24}, and enhancing the error correction capabilities for existing ones \cite{TBF18,BonillaAtaides2021,PBSD21,HBTK23}.

Most decoding algorithms make simplified assumptions about the underlying error model, potentially reducing the performance of quantum error correction. Ideally, we aim to optimize decoding to the experimentally measurable characteristics of the specific noise process affecting our physical system.
Some recent work has taken this direction. In Refs.~\cite{TBF18,HNB20,LPH21,ALMS22,XMSK23,KKK23}, surface code decoders have been adapted to an inherent bias \cite{PJGG20,BonillaAtaides2021,SJCT23} between the probabilities of $X$ and $Z$ type errors and to the correlated errors in Ref.~\cite{BLSW22} observed in some hardware devices. A bias is only one of the exponentially many features that describe the noisy dynamics of a quantum system. In some cases, the decoders have been adapted to specific instances of non-Pauli noise models, including rotation errors \cite{BEKP18,CWBL17,GSVB13}, correlated errors \cite{NB19}, amplitude damping noise \cite{JLNM22}, erasure \cite{KHV23,SJC23,wu2022erasure}, leakage \cite{PB07,AF13,miao2023overcoming}, and circuit-level noise for fault tolerance \cite{CP18,BVM23}. In general, fully adapting a decoding algorithm to the microscopic details of an underlying
physical noise process is achieved by maximum likelihood (ML) decoders \cite{IP15,HL11,JKH20,BSV14,Yoder23,P20}. Logical error suppression with an ML decoder indicates the best achievable performance for a code and error-model pair. However, estimating this quantity presents challenges, depending on the family of codes. Firstly, ML decoding is computationally intractable for general stabilizer codes; concatenated codes are among the few families of quantum error-correcting codes known to have an efficient ML decoder~\cite{P06}. Secondly, an ML decoder requires precise knowledge of the underlying physical error model that normally requires an exponential number $\cO(4^{n})$ of parameters to represent and an exponential number of experiments to learn. Unfortunately, process tomography methods to fully characterize an error process  are neither scalable nor generally accurate~\cite{QPTforQFT, SRMG17,LLC24}. The problem of inefficient error-representation can be overcome with recursive techniques for concatenated codes, and the problem of exponential experimental overhead  can be overcome thanks to the discovery of efficient error learning techniques \cite{ESMR07, EWPM19,CDHO23}, both of which are described below. 

In this paper, we introduce and demonstrate a method that achieves a significant enhancement in the performance of quantum error correction through a decoding strategy informed by error model information obtained through efficient state-of-the-art error-learning experiments. We show that the suppression of logical error rates consistently improves with the amount of noise characterization data available, achieving an order-of-magnitude improvement even when the calibration data include only a small fraction of the total parameters required to fully characterize all the Pauli error rates in the underlying noise process. Our findings rely on a class of scalable Pauli error learning techniques based on the Pauli eigenvalue estimation ideas of Emerson \textit{et al.}~\cite{ESMR07} and the cycle benchmarking protocol of Erhard \textit{et al.}~\cite{EWPM19}, which culminate in the analysis and robustness guarantees of the cycle error reconstruction (CER) protocol implemented in Ref.~\cite{TrueQ20} and described in Ref.~\cite{CDHO23}. The CER protocol enables efficient estimation of a subset of the Pauli error rates of an error channel to within multiplicative precision. The feasibility of these error-learning methods has already been  experimentally demonstrated across several platforms~\cite{ESMR07,EWPM19,Hashim2021,CDHO23,TrueQ20}, and most recently in  Ref.~\cite{fazio2025characterizingphysicallogicalerrors} in which this approach was applied to learn a subset of the Pauli error rates on the physical qubits that was small enough to be scalable but large enough to reliably infer the resulting logical error rate in a transversal CNOT gate implemented on a 16-qubit trapped-ion quantum computer. Similarly, our proposed method does not rely on learning all of the (exponentially many) parameters of the error channel, but requires only a very limited subset of Pauli error rate data characterizing the $K-$largest Pauli error rates in a noise process, where $K\ll 4^{n}$. The practical relevance of this input data set can be justified on physical grounds by post-processing ideas that extend Pauli eigenvalue estimation \cite{ESMR07} and cycle benchmarking \cite{EWPM19} methods to show that the $K-$largest Pauli error rates on $n$ physical qubits using a number of experiments proportional to $K$ given certain strong but plausible assumptions on the (otherwise unknown) error model~\cite{FW20,HYF21}. 

Our methodology is twofold. First, we design a heuristic technique to reconstruct a complete set of Pauli error probabilities from limited error characterization data. Second, we employ an ML decoder using the reconstructed Pauli error distribution. Our tools are designed for the family of concatenated quantum codes \cite{K96,JL14}, which are among the few families of quantum error-correcting codes known to have an efficient ML decoder \cite{P06}.

The paper is organized as follows. \Cref{sec:background} provides the necessary background and essential mathematical foundations to lay the foundation for the main contribution of this paper, which is detailed in \cref{sec:uss}. In \cref{sec:results}, we demonstrate the efficacy of our approach by numerically estimating the reduction in logical error rates achieved for concatenated Steane codes under various realistic error models. These findings highlight the potential of cleverly using a small portion of the CER data to significantly enhance performance, surpassing the results obtained without adapting the decoding techniques to the underlying error model. Finally, in \cref{sec:conclusion}, we provide concluding remarks and outline interesting future directions.

\section{Background} \label{sec:background}
This section provides the technical background necessary to obtain a thorough understanding of our results. It is organized as follows. We begin with a brief introduction to the mathematical modeling of physical noise processes. Then, in \cref{sec:cer_nr}, we discuss an important method of estimating the probabilities of Pauli errors in an error model, which sets the context for the research problem we want to address. Finally, in \cref{sec:qec}, we recall some basic knowledge of quantum error correction for concatenated codes using the stabilizer formalism.

\subsection{Physical error processes} \label{sec:noise}
Markovian processes encompass a wide range of physically motivated noise mechanisms. Typically, Markovian noise arises from a system's interactions with its environment and is mathematically described by Completely Positive Trace Preserving (CPTP) maps \cite{C75,Kraus83,CN97,L19}. These maps, denoted as $\cE$, transform any quantum state $\rho$ into its \emph{noisy} counterpart, $\cE(\rho)$.

An interesting case of CPTP maps is unitary errors:
\begin{gather}
\cE_{\cU}(\rho) = U ~\rho~ U^{\dagger} ~ .  \label{eq:unitary_channel}
\end{gather}
Another popular class of CPTP maps is \emph{Pauli channels} where the noisy state $\cE(\rho)$ is described as a mixture of any of the Pauli matrices applied to $\rho$:
\begin{gather}
\cE_{\cP}(\rho) = \sum_{P \in \cP_{n}} \Prob(P) ~ P ~ \rho ~ P ~ , \label{eq:pauli_channel}
\end{gather}
Coherent errors and Pauli errors constitute two extremes of the noise spectrum described by CPTP maps.

We can describe the effect of a general CPTP map as:
\begin{gather}
\cE(\rho) = \sum_{P,Q \in \cP_{n}}\chi_{P,Q}(\cE) ~ P ~ \rho ~ Q ~ , \label{eq:chi}
\end{gather}
where $\chi_{P,Q}(\cE)$ are complex numbers that correspond to a unique pair of $n-$qubit Pauli matrices $P, Q \in \cP_{n}$. The Hermitian matrix $\chi(\cE)$ with unit trace is known as the $\chi-$matrix representation of $\cE$ \cite{YK05,GDW09,WBJC15}. We will often refer to the diagonal entries of the $chi-$matrix, $\{\chi_{P,P}(\cE)\}_{P\in\cP_{n}}$, as Pauli error rates. For a Pauli channel in \cref{eq:pauli_channel}, $\chi_{P,P}$ denotes the probability of a Pauli error $P$.

Various metrics exist to quantify the impact of noise described by a CPTP map \cite{GB15,Watrous_2018,KLDF16,IP17}. Among these, average fidelity, denoted by $F_{g}$, is one of the most common choices:
\begin{gather}
F_{g} = \int\mathrm{d}\psi ~ \tr(~ |\psi\rangle\langle\psi| \cdot \cE(|\psi\rangle\langle\psi|) ~ ) ~ , \label{eq:avg_fidelity}
\end{gather}
particularly because there are efficient experimental protocols to estimate it. We will use a closely related quantity, denoted by $\epsilon(\cE)$, referred to as the \emph{process infidelity} in \cite{EWPM19,CAE19,B21,PYBA24}, to quantify the \emph{physical error rate}:
\begin{gather}
\epsilon(\cE) = \dfrac{3}{2}(1 - F_{g}) \label{eq:pro_infid} ~ .
\end{gather}
It can be shown that $\epsilon(\cE) = 1 - \chi_{0,0}(\cE)$ \cite{JK11,I18}.

\subsection{Error Learning and Cycle Error Reconstruction} \label{sec:cer_nr}
An efficient noise characterization approach should focus on acquiring (only) the key parameters necessary to understand the critical effects of the underlying noise process on fault-tolerance schemes. One such important parameter is the average fidelity of a noise process, denoted by $\epsilon(\cE)$, which can be efficiently estimated using Randomized Benchmarking (RB) experiments \cite{emerson2005scalable,PhysRevA.80.012304,MGE11,MGE12}. While RB experiments provide information about the fidelity averaged over a set of Clifford operations, learning more fine-grained error information is critical to optimizing control and predicting error correction performance. In particular, it was critical to devise methods to learn (i) more fine-grained error properties than just the fidelity $\epsilon(\cE)$, which is only one of the $\cO(4^{2n})$ parameters that completely specify an $n-$qubit physical noise process $\cE$ and (ii) the error properties associated with specific cycles of interest (e.g., parallel sets of gate operations) rather than an error map averaged over a set of Clifford operations. 

The solution to (i) was developed already in 2007 by Emerson \textit{et al.}~\cite{ESMR07}, which demonstrated that symmetrizing (or twirling) a quantum channel, e.g., an error channel or an `error oracle', via $\mathcal{P}_n$  simplifies the channel to a \emph{diagonal} Pauli-transfer matrix (PTM), i.e., a Pauli channel. Moreover, the individual eigenvalues of that channel can then be efficiently learned (ie, experimentally measured) by propagating a Pauli operator through the twirled channel. This enables a protocol for learning the eigenvalues of the error channel. Moreover, the full set of Pauli operators can be generated by applying  local Clifford operations drawn from $\mathcal{C}_1^{\otimes n}$ to the $|0\rangle^{\otimes{n}}$ state and applying their inverse to the output state. When the local Clifford operators are sampled randomly, this is equivalent to a local Clifford twirl of the error channel, which creates degeneracies in the PTM eigenvalues. Ref.~\cite{ESMR07} further demonstrated that the error probabilities of the PTM can related to the learned eigenvalues via an invertible linear transformation, and showed that \emph{efficient} learning of error probabilities of the PTM is possible via post-processing by making reasonable assumptions on the correlations present in the error model. Ref.~\cite{ESMR07}  showed, as an example, that efficient learning of all low-weight error probabilities is possible assuming permutation symmetry over the qubits, which is equivalent to measuring only the Hamming weight of the output strings under the Pauli-channel learning experiment. These ideas were then demonstrated experimentally on a 3-qubit NMR system~\cite{ESMR07}.

The solution to (ii) was developed in the cycle benchmarking protocol of Erhard \textit{et al.} \cite{EWPM19} by generalizing the error amplification ideas of RB to the setting of Pauli channel learning. In particular, Ref.~\cite{EWPM19} showed that by repeating a fixed cycle of interest under randomized compiling \cite{WE16}, one can apply the methods of Ref.~\cite{ESMR07} to learn the eigenvalues of the error channel \emph{associated with a specific cycle of interest}. In this way the cycle benchmarking method creates Pauli channel in place of the memory channel, or `oracle' assumed in the Pauli channel learning protocol of Emerson \textit{et al.}~\cite{ESMR07}. As a result, the eigenvalues and error probabilities of the error channel of a specific cycle can be learned to multiplicative precision in a SPAM-free way. This protocol was then 
 demonstrated experimentally on a 10 qubit ion trap~\cite{EWPM19}. 

These ideas have been combined into the Cycle Benchmarking technique known as Cycle Error Reconstruction (CER) \cite{CDHO23}: an experimental protocol to measure the individual $4^{n}$ Pauli eigenvalues and Pauli error rates (of an error map $\cE$ associated with any cycle or any circuit block of interest) to within multiplicative precision and with rigorous robustness guarantees. \Cref{sec:app_cer_nr} provides an overview of CER. These protocols are practical and efficient and have been  demonstrated experimentally on multiple small-to-medium scale quantum computing platforms ~\cite{ESMR07,EWPM19,Hashim2021,CDHO23,TrueQ20,fazio2025characterizingphysicallogicalerrors}.
Subsequent work~\cite{FW20,HYF21} leverages the solutions to (i) and (ii) and adds additional post-processing steps to justify, under  assumptions on the allowed error model correlations,  direct estimation of the $K-$largest Pauli error rates of an error channel $\cE$ using a number of experiments that scale linearly with $K$ \cite{HYF21}.

In this work we apply and extend  the above ideas with a focus on enabling the efficient learning the errors that are relevant in the context of enhancing decoding for quantum error correction.

\subsection{Quantum error correction} \label{sec:qec}
In this section, we will outline the essential concepts of quantum error correction using concatenated codes within the stabilizer formalism. Our focus will lie on correcting errors described by a general CPTP map, especially including non-Pauli error models. For a more elaborate explanation, see \cref{sec:qec_non_Pauli} and the references therein.

A $[[n,k,d]]$ stabilizer code \cite{GotPhD97} is given by the common eigenspace with an eigenvalue $+1$ of $n-k$ commuting operators $S_{1}, \ldots, S_{n-k}$, called stabilizer generators. The group $\cS = \langle S_{1}, \ldots, S_{n-k}\rangle$ is called the stabilizer group. The encoded version of a $k-$qubit state is the $n-$qubit state denoted by $\ol{\rho}$. For the results in this paper, we will limit ourselves to codes that encode only one logical qubit, i.e. $k=1$.

An $[[n,k]]$ stabilizer group can be used to derive a canonical choice of a basis for the $n-$qubit Pauli group. To start with we have $n-k$ stabilizer generators for $\cS$. We define the group of logical operators for a code, denoted by $\cL = \cN(\cS)/\cS$, where $\cN(\cS)$ is the normalizer of the stabilizer group in the Pauli group. Note that $\cL$ consists of $2k$ generators. The remaining $n-k$ generators, labeled $T_{1}, \ldots, T_{n-k}$, form the group of \emph{Pure Errors} \cite{PP13,IP15,Beale23}, denoted by $\cT$, where $\cT = \cN(L)/\cS$. Consequently, an $n-$qubit Pauli error $E$ can be written as a product of three terms:
\begin{gather}
E = T\cdot L\cdot S \label{eq:tls}
\end{gather}
where $S\in \cS$, $L \in \cL$, and $T\in\cT$.

QEC in the stabilizer formalism is a two-fold procedure. The first step includes measuring each of the stabilizer generators on the noisy state $\cE(\ol{\rho})$. The outcomes of the measurement form the error syndrome, denoted by $\vec{s}$.
While $T$ is entirely fixed by the syndrome $\vec{s}$: $T \equiv T_{s}$, the problem of determining the most likely logical operator, denoted by $L_{s}$, is called \emph{Maximum Likelihood (ML) Decoding}:
\begin{gather*}
L_{\vec{s}} = \argmax{L\in\cL}~\Prob(L ~|~ s) \label{eq:ml_decoder} ~.
\end{gather*}
where
the probability of a logical operator, $\Prob(L | s)$, is given by
\begin{gather*}
\Prob(L | s) = \sum_{S \in \cS}\Prob(T_{s}\cdot L\cdot S) ~ . \label{eq:prob_Ls}
\end{gather*}

When selecting a decoder for a QEC code, one must balance efficiency and accuracy. Although ML decoding is theoretically optimal, it remains computationally intractable \cite{IP15}. In contrast, \emph{Minimum Weight (MW)} decoding is an alternate suboptimal strategy with lower thresholds but offers high efficiency, as it can be implemented using a lookup table \cite{FAMJ12,Google25}.

With a slight loss of generality, we assume that syndrome measurement and the application of the recovery operation prescribed by the decoder $L_{s}$ are executed perfectly. Under this perfect QEC assumption, the combined effect of noise and QEC preserves the codespace. We refer to this composite process as an effective channel and characterize it by its direct action on encoded states, described by a $k-$qubit CPTP map, denoted as $\cE^{s}_{1}$. The \emph{average effective channel}, $\ol{\cE}_{1}$, is then given by
\begin{gather*}
\ol{\cE}_{1} = \sum_{\vec{s} \in \bZ_{2}^{n-k}} \Prob(\vec{s})~\cE^{s}_{1}  ~ ,
\end{gather*}
which quantifies the average strength of noise that remains in the encoded information after QEC. We define \emph{logical error rate} as the infidelity of $\ol{\cE}_{1}$.

Concatenation is a well-established technique for constructing a family of codes with increasing sizes that exhibit a threshold \cite{KL96,GotPhD97,YTY24,YK24}. This approach involves recursively encoding the physical qubits of a quantum error-correcting (QEC) code into another QEC code. Repeating this process $\ell$ times results in the \emph{level$-\ell$ concatenated code}. For simplicity, we consider the same code at each level. Appendix~\ref{sec:decoding_concatenated} outlines the computation of the average effective channel $\ol{\cE}_{\ell}$, which characterizes the combined impact of noise and QEC on the level$-\ell$ concatenated code.

\section{Working with partial noise characterization information} \label{sec:uss}
Recall that a maximum likelihood decoder defined in \cref{eq:MLD_prob_L} requires complete knowledge of the error model, namely the $4^{n}$ Pauli error rates for a $[[n,k,d]]$ quantum error correcting code. However, in a practical scenario, we assume that only a handful of these correspond to the $K-$largest Pauli error rates estimated using techniques developed in \cite{HYF21}, where $K \ll 4^{n}$. In what follows, we will use $\cN_{K}$ to denote the set of Pauli errors that correspond to the $K-$largest Pauli error rates, and $\chi_{K}$ to refer to the respective error rates, and $\cD_{K}$ to denote a decoder that computes a recovery operation conditioned on an error syndrome $s$, given $\chi_{K}$. 

The limited data, constituting only $K$ Pauli error rates, is insufficient to directly employ an ML decoder. We need to first complete the Pauli error distribution by assigning probabilities to those errors in $\cP_{n}$ that are not in $\cN_{K}$. Hence, given an error syndrome, we design $\cD_{K}$ to apply a twofold strategy:
\begin{itemize}
\item[(i)] Apply a heuristic algorithm to obtain the error rates of all $4^{n}$ Pauli error rates from $\cN_{K}$, affecting a code block.
\item[(ii)] Apply an ML decoding strategy given the error syndrome and the heuristically reconstructed Pauli error distribution from step (i).
\end{itemize}

A special case of the limited error characterization data is where CER is used to estimate only the probabilities of the single-qubit errors are known. A common assumption used to infer the probabilities of higher-weight errors is that errors are independently and identically distributed across qubits. Under this assumption, the probability of a high-weight error is determined by the product of the probabilities of the single-qubit errors that form the tensor product representation of the multi-qubit error. Inspired by this approach, we design a greedy heuristic algorithm that leverages the probabilities of Pauli errors in $\cN_{K}$ to estimate the distribution of all $4^{n}$ Pauli errors affecting the code. Our heuristic algorithm, which we call \emph{Uncorrelated Split Search}, abbreviated as \emph{USS}, operates on the following principle. Consider an $n$-qubit Pauli error $P$ that is not included in $\cN_{K}$. Note that $P$ can be expressed as $P = P_{1} \cdot P_{2}$ for various choices of Pauli errors $P_{1}$ and $P_{2}$ acting on disjoint subspaces. In essence, $P_{1}$ acts on a subspace of dimension $2^{m}$, while $P_{2}$ acts on a subspace of dimension $2^{(n-m)}$. For each of these decompositions, our estimate of the probability of $P$ is merely the product of the probabilities of $P_{1}$ and $P_{2}$: $\Prob(P) = \Prob(P_{1})\Prob(P_{2})$. Each bi-partition of the $n$ physical qubits into sets $B_{1}$ and $B_{2}$ corresponds to an error $P_{1} \equiv P_{1}(B_{1})$ with nontrivial support in $B_{1}$ and another error $P_{2} \equiv P_{2}(B_{2})$ with non-trivial support in $B_{2}$. Denoting the set of bi-partitions of $n$ qubits as $\cB$, the overall probability of the error $P$ adopts a straightforward sum-product structure:
\begin{gather}
\Prob(P) = \sum_{{B_{1}, B_{2}} \in \cB(P)}\Prob(P_{1}(B_{1}))\Prob(P_{2}(B_{2})) ~ . \label{eq:prob_P_P1_P2}
\end{gather}
Physically speaking, we are assuming that the error $P$ can occur whenever both $P_{1}$ and $P_{2}$ occur. It should be noted that there are several ways to combine two $n-$ qubit errors to produce $P$. However, our heuristic ignores the possibilities where the errors in the decomposition have overlapping supports. Since $P_{1}$ and $P_{2}$ act on fewer than $n$ qubits, in a physically plausible error model, we are more likely to find the error probabilities of $P_{1}$ and $P_{2}$ in the CER data provided $\cN_{K}$. If they are not found, we recursively apply the aforementioned strategy to both $P_{1}$ and $P_{2}$ until their probabilities are found in $\cN_{K}$.

While it is reasonable to assume that $\cN_{K}$ contains all single-qubit errors, this is not always the case when $K < (3n+1)$ or if the physical noise model exhibits strong correlations. In such scenarios, we might encounter a single-qubit error $P_{1}$ whose probability cannot be inferred from $\cN_{K}$. To address this, we establish the base case where $K = 0$, corresponding to $\cN_{0} = {I}$. This implies that we always know the average physical error rate $\epsilon(\cE)$. Based on this assumption, we will use a simple ansatz to assign the probability of any single qubit error $P$:
\begin{gather}
\Prob(P) = \dfrac{1 - (1 - \epsilon)^{1/n}}{3} (1 - \epsilon^{1/n})^{n-1} ~ , \label{eq:prob_single_qubit_depolarizing}
\end{gather}
where $(1-\epsilon)^{1/n}$ is the fidelity of a single qubit depolarizing channel, assuming that $\cE$ is an i.i.d depolarizing channel with average physical error rate $\epsilon(\cE)$. Now, we are ready to describe the formal algorithm of the \emph{USS} method stated in \cref{alg:uss}.

\begin{algorithm}[H]
\caption{The following pseudo-code describes the \emph{Uncorrelated Split Search} (abbreviated as \emph{USS}) strategy, a greedy algorithm employed to estimate the probabilities of all $4^{n}$ Pauli errors, leveraging the probabilities of $K$ Pauli errors from CER. We will call the below function with inputs $\cN_{K}, \chi_{K}$. Upon completion, $\hat{\cN}$ will include all $4^{n}$ Pauli errors, and $\hat{\chi}_{K}$, their respective probabilities which will serve as input to the ML decoder.}
\label{alg:uss}
\begin{algorithmic}[1]
\Procedure{Uncorrelated Split Search}{$P, \hat{\cN}, \hat{\chi}_{K}$}
\If{$P$ is found in $\hat{\cN}$}
\State \Return $\chi_{P,P}(\cE)$
\ElsIf{$P$ is a single qubit error}
\State $\epsilon_{0} \gets 1 - (1 - \chi_{I,I})^{1/n}$
\State $p_{\star} \gets \dfrac{\epsilon_{0}}{3}(1-\epsilon_{0})^{n - 1}$ 
\Else
\State $p_{\star} \gets 0$
\For{$B_{1}, B_{2} \in \cB(P)$}
\State $p_{1} \gets $ \Call{Uncorr Split Search}{$P_{1}(B_{1}), \hat{\cN}, \hat{\chi}_{K}$}
\State $p_{2} \gets $ \Call{Uncorr Split Search}{$P_{2}(B_{2}), \hat{\cN}, \hat{\chi}_{K}$}
\State $p_{\star} \gets p_{\star}~ +~ p_{1} ~ p_{2}$
\EndFor
\EndIf
\State Add the error $P_{i}$ and its probability $p_{\star}$ to $\hat{\cN}$.
\State \Return $p_{\star}$
\EndProcedure
\end{algorithmic}
\end{algorithm}

It is important to note that every Pauli error $P\in\cP_{n}$ is only visited once when it is assigned a probability. Once its probability is assigned, it is added to the set $\hat{\cN}$. After this, its probability is simply looked up from $\hat{\cN}$ whenever needed. At the end of the algorithm, the set $\hat{\chi}_{K}$ should contain all the $4^{n}$ Pauli error probabilities. Hence, the time complexity of our \emph{USS} method in deriving the entire Pauli error distribution is linear in the number of Pauli errors.

Note that the \emph{USS} technique can be applied to any dataset containing the $K \leq 4^{n}$ Pauli error rates. The authors in \cite{FW20} present general methods for estimating error rates for Pauli errors from an arbitrary set, such as low-weight Pauli errors. Subsequently, high-weight errors are estimated by assuming a correlation length within the system.

\section{Results} \label{sec:numerics}
Our goal is to demonstrate the importance of CER data and the \emph{Uncorrelated Split Search} technique in improving the performance of the ML decoder for concatenated codes. We achieve this by presenting numerical evidence across a broad range of realistic error models. In particular, we investigate a case where only the largest $K$ Pauli error rates are extracted using CER. Our aim is to prove that our methodologies are widely applicable to various coherent and incoherent errors characterized by CPTP maps, going beyond error models that can be closely approximated by Pauli errors.


Our numerical analysis focuses on concatenated Steane codes; however, our conclusions remain qualitatively similar for other families of concatenated codes as well. We will use $n$ to denote the size of a code block in the concatenated code; $n = 7$ for the Steane code. The total number of physical qubits in the level$-\ell$ concatenated Steane code is $n^{\ell}$. So, a realistic error model for the concatenated Steane code is described by a CPTP map on $n^{\ell}$ qubits. Recall that the $n^{\ell}$ physical qubits of a level$-\ell$ concatenated code belong to $n^{\ell-1}$ disjoint code blocks. Since the encoding circuit of the concatenated code shown in \cref{fig:schematic_encoder_concatenated} resembles a tree where every layer is an identical encoding circuit, we assume that noise acts in an independent and identical fashion on all the $n^{\ell-1}$ code blocks of the concatenated code \cite{RDM02,CWBL17,IP17}.

\subsection{Error models} \label{sec:unitary_error_models}
We want to numerically investigate the performance gain in QEC achieved by leveraging a small fraction of the CER data and the USS algorithm across a broad class of physically motivated error models. While the independent and identically distributed (i.i.d.) Pauli error model facilitates code analysis and design, extending beyond the Pauli framework is essential for accurately capturing realistic noise \cite{TS14,DP17,RK21,Beale23}. In most experimental implementations, noise is characterized by a quantum channel that exhibits spatial correlations and varies between qubits. Additionally, noise correlations arise due to the engineering challenge of addressing individual qubits without inadvertently affecting their neighbors \cite{CTBS21,NB19,SPRY20,RHNH21,H20}.

\begin{figure}
\begin{center}
\includegraphics[scale=0.28]{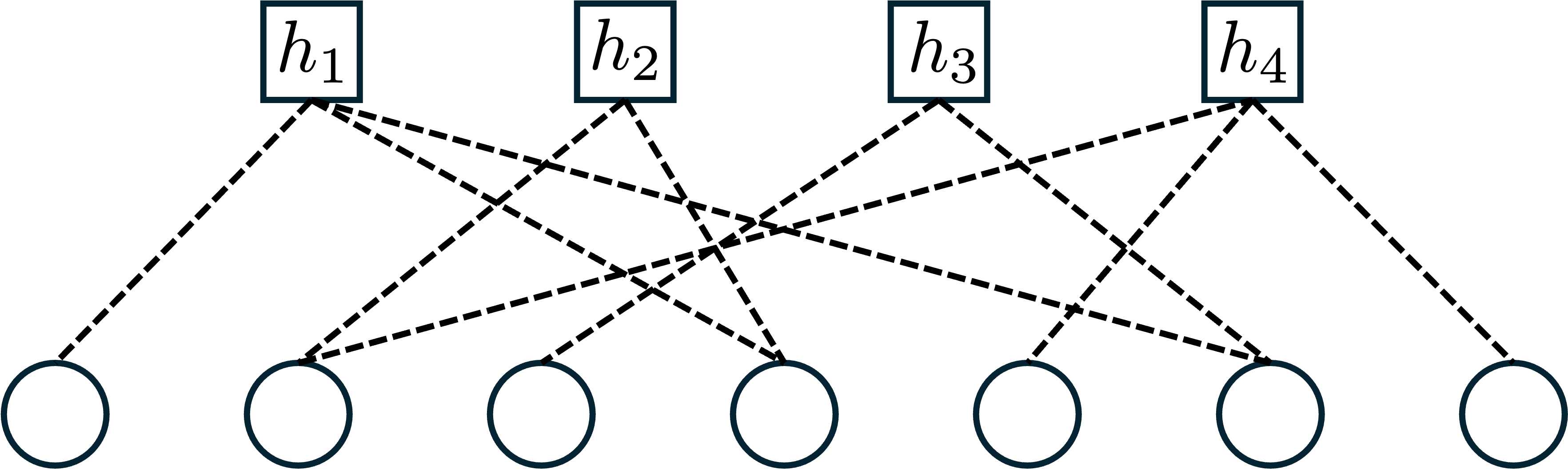}
\caption{Figure showing a factor graph used for constructing an error in the CG1D error model. Each square node specifies a local interaction supported on the qubits of a Steane code block, represented by circular notes that are incident to respective the square node. In the example shown, the local interactions denoted by $h_{1}, h_{2}, h_{3}$ and $h_{4}$ are supported on qubits $\{1,4,6\}$, $\{2,4\}$, $\{3,6\}$, and $\{2,6,7\}$ respectively. The sum of these local interactions is a $n-$qubit Hamiltonian denoted by $H$ which is used to construct a unitary error $U = \exp\left(-i~t~\sum_{i=1}^{4}h_{i}\right)$.}
\label{fig:factor_graph}
\end{center}
\end{figure}

A broad class of realistic noise processes can be physically modeled by considering qubit interactions governed by a Hamiltonian $H$ \cite{VNHG19,BW24,A06,CW20}. Of particular interest are Hamiltonians expressed as a sum of local interactions: $H = \sum_i h_{i}$, where $h_{i}$ are random Hermitian matrices of dimension $2^{m}$ for $m\leq n$.  Such local interactions are characteristic of Hamiltonians commonly studied in many-body physics. These error models, referred to as \emph{Coarse-Grained 1D} (CG1D) error models in \cite{FW20,HF23}, are effectively represented by the factor graph in \cref{fig:factor_graph}, which illustrates the support of these local interactions. In particular, CG1D models have been shown to provide an accurate description of the physical noise processes that affect certain Google devices \cite{HF23}.

We perform numerical studies on random instances of the CG1D error model by generating random Hamiltonians describing interactions between the qubits of a Steane code block in a concatenated code. In order to construct a random Hamiltonian, we sample local Hamiltonians supported on a subset of the seven qubits in the Steane code block. We choose the sizes of these subsets from a Poisson distribution whose mean is fixed to reflect a mean correlated length $\lambda$ in a system. By setting $\lambda$ to be greater than one, we can sample a large number of local interactions that involve multiple qubits.

\begin{figure*}
\begin{center}
\subfloat[\label{subfig:cg1d_a}]{%
  \includegraphics[scale=0.12]{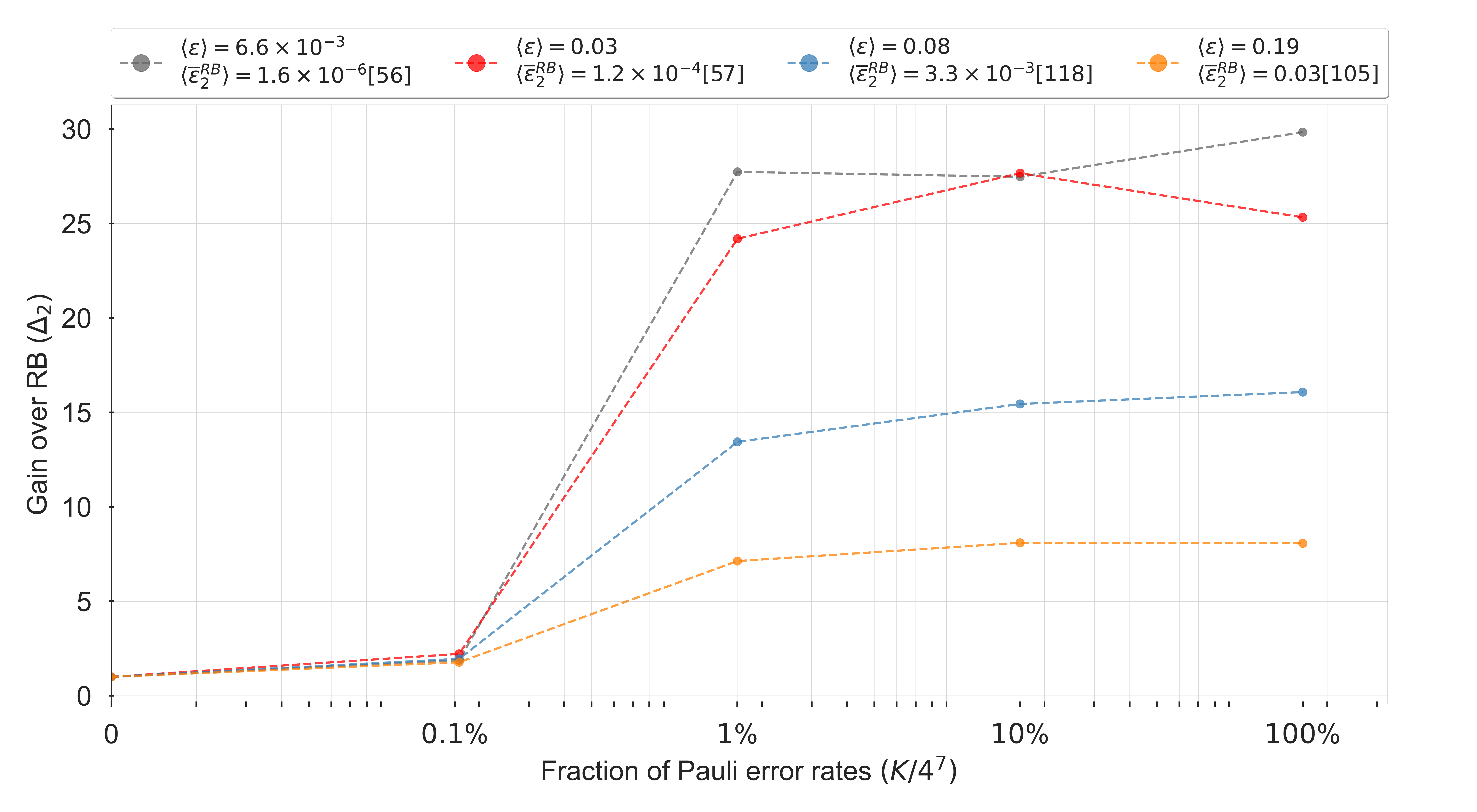}%
}

\subfloat[\label{subfig:cg1d_b}]{%
  \includegraphics[scale=0.12]{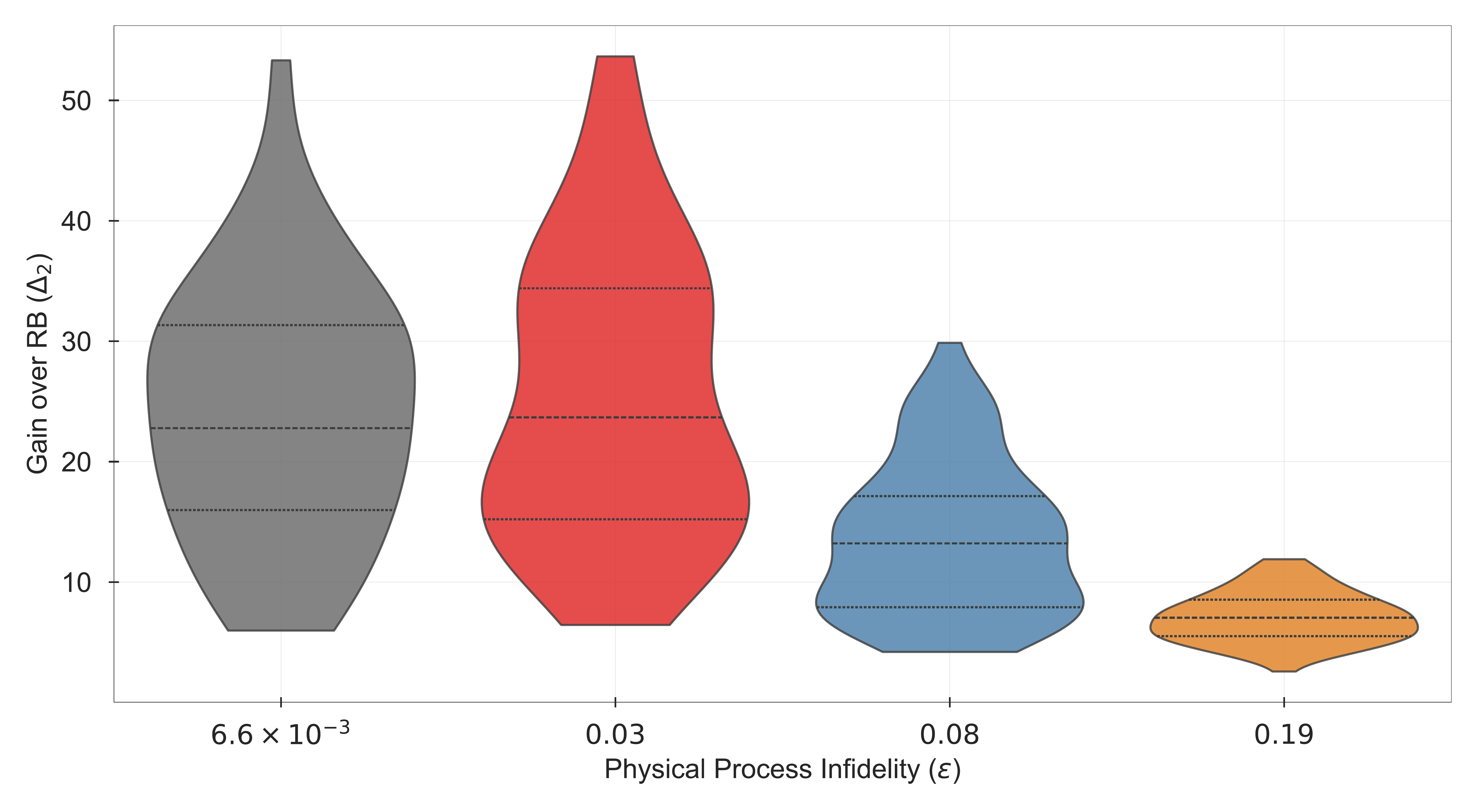}%
}

\caption{Figures illustrating the performance enhancement in the QEC capabilities of the Steane code obtained by using partial \emph{error characterization data using CER} and employing the \emph{Uncorrelated Split Search} (\emph{USS}) method outlined in \cref{alg:uss} as a function of the number of Pauli error rates estimated from CER. This enhancement is quantified by $\Delta_{2} = \ol{r}_{2}(\cE ~ ; ~ \cD_{1})/\ol{r}_{2}(\cE ~ ; ~ \cD_{K})$. \Cref{subfig:cg1d_a} presents $\Delta_{2}$ based on numerical simulations of QEC under random unitary noise processes arising from the CG1D error model. These random unitary errors are grouped into bins with similar physical infidelities, and the median performance gain for each bin is shown. Each bin is distinguished by a unique color, as indicated in the legend, which also details the number of errors in each bin and their average physical and logical infidelities. In all but the extremely high-noise regime, the ML decoder achieves a performance gain of approximately $10X$ when it assumes knowledge of merely $\approx 1\%$ of the available error characterization data from CER, specifically $163$ Pauli error rates. \Cref{subfig:cg1d_b} illustrates the variation in performance gains across individual physical noise processes within the four bins considered in \cref{subfig:cg1d_a}, when the ML decoder has access those 163 Pauli error rates that constitute the $1\%$ of the error data from CER. The width of a violin-shaped region at a specific height ($\Delta_{2}$) is proportional to the number of noise-processes that exhibited the corresponding performance gain. The horizontal dashed lines represent the first quartile, median, and third quartile of the performance gains, respectively. While the majority of performance gains fall between $15X$ and $30X$, gains in the low-noise regime can reach as high as $50X$.}\label{fig:cg1d}
\end{center}
\end{figure*}

\subsection{Quantitative estimation of the performance gain}
We want to demonstrate a performance improvement with the ML decoder under a variety of noise models described in the earlier section, with the amount of error characterization data from CER supplied to the decoder. However, first, we want to establish a baseline for comparison by defining a reference scenario representing the performance of a decoder $\cD_{1}$, which only has access to the average fidelity of the underlying physical noise process. Consequently, for any given physical noise process, we intend to compare two metrics: (i) the performance of a decoder $\cD_{K}$ for $K > 0$ against (ii) the performance of $\cD_{1}$. We gauge this comparison qualitatively using $\Delta_{\ell}$ for a level$-\ell$ concatenated code which is defined as
\begin{gather}
\Delta_{\ell}(K) = \dfrac{\ol{r}_{\ell}(\cE ~ ; ~ \cD_{1})}{\ol{r}_{\ell}(\cE ~ ; ~ \cD_{K})} \label{eq:def_delta} ~ ,
\end{gather}
where we have used $\ol{r}_{\ell}(\cE ~ ; ~ \cD_{K})$ to denote the average logical error rate for the level$-\ell$ concatenated code where $\cE$ is the physical noise process and $\cD_{K}$ is the decoder used. In what follows we will refer to $\Delta_{\ell}(K)$ as \emph{gain}\footnote{We recognize that an alternative reference for comparison, in place of the performance of $\cD_{1}$, is the performance of the Minimum Weight Decoder (MWD), denoted $\cD_{MWD}$. Despite its widespread use \cite{CWBL17,FAMJ12}, the MWD is generally considered suboptimal compared to $\cD_{1}$ because it accounts for the most likely error instead of the most likely logical coset \cite{YL22,dFOC24}. Consequently, our estimation of gain is a conservative one. Specifically, when we set the numerator in \cref{eq:def_delta} to $\cD_{MWD}$, we observe that the gain $\Delta_{\ell}$ is several orders of magnitude higher than the values reported in our estimates.}.

Note that a gain larger than one signifies a relative performance gain that is achieved by the ML decoder that has access to $K$ Pauli error rates in $\cN_{K}$, in conjunction with the \emph{USS} method in \cref{alg:uss}. Across a range of physically relevant error models, we consistently observe a gain exceeding one, that consistently improves with the fraction of accessible CER information by the decoder. Notably, even when only $1\%$ of the total CER information is available to the decoder, the observed gain is approximately $10$.

Note that the gain defined in \cref{eq:def_delta} strictly depends on the microscopic details of the underlying physical noise process $\cE$. However, for a broader analysis of typical performance improvements across a range of random instances of $\cE$, we examine the gain within an ensemble of random physical noise processes having similar levels of physical noise strength, quantified by their error rates. We categorize these physical noise channels based on their error rates falling within specific intervals, forming what we call a ``bin'': $\mathfrak{B} = {\cE ~ : ~ b_{1} \leq r(\cE) < b_{2}}$. For each bin, we report the median gain observed over all physical noise processes in the bin, since it reflects the most frequently observed gain in the corresponding ensemble of physical noise processes. In our graphical presentations, we distinguish bins with different colors and provide the average physical as well as logical error rate of the physical noise processes in each bin, along with the size of the bin.

\subsection{Performance analysis} \label{sec:results}
\Cref{fig:cg1d} shows gains for the level$-2$ concatenated Steane code achieved using an ML decoder that leverages Pauli error rates in $\cN(K)$ along with the \emph{USS} method for several values of $K$. The trends in the figure are drawn from an ensemble of $336$ Unitary errors drawn from a CG1D error model. Each unitary error is constructed by exponentiating a Hamiltonian multiplied with a time parameter $t$, chosen between $0.001$ and $0.1$. The Hamiltonian is composed of random local Hamiltonians that are supported on random subsets of qubits whose sizes are sampled from the Poisson distribution with mean two. The mean is chosen to be two in order to introduce a higher frequency of many-body interaction terms in the $n-$qubit Hamiltonian.

All of the random unitary errors are organized into bins according to their physical infidelities. We used a distinct color to identify a bin and provided the average physical as well as logical error rate corresponding to the unitary errors in each bin and the size of the bin in the legend. The performance gain reported in the Y-axis of \cref{fig:cg1d} is the median of the gains associated to the unitary errors in the respective bins, while the error bars denote the first quartiles, indicating that there are instances of physical noise processes where the performance gain is substantially higher than the median. We observe that even with as low as $1\%$ of all Pauli error rates within a code block, i.e., $K = 163$, we are able to achieve a typical performance gain of $10X$ over $\cD_{1}$ which only has access to the average physical error rate.

While \cref{fig:cg1d} highlights the performance gains achieved under CG1D error models, we have observed similar improvements across a broad range of noise models. These include coherent errors from random unitary circuits with one- and two-qubit gates, as well as Markovian noise represented by $n$-qubit CPTP maps, which are constructed by composing multiple one- and two-qubit quantum channels. A detailed discussion of these results can be found in the appendix sections \ref{sec:random_unitary} and \ref{sec:cptp_error_model}. The outstanding observation across all the error models considered is a $10X$ performance gain achieved with just $1\%$ of the available Pauli error rates in the CER data—specifically, the 164 largest Pauli error rates. These results underscore the versatility of CER data and the effectiveness of the \emph{USS} in QEC for general Markovian noise.

The performance improvements depicted in \cref{fig:cg1d} are attributed to the combined impact of the partial CER data regarding the underlying noise process and the \emph{USS} algorithm (\cref{alg:uss}). In the following analysis, we aim to isolate the individual contributions of the partial CER data and the \emph{USS} algorithm to these gains. For an $n$-qubit noise process $\mathcal{E}$, we introduce three key notations: (i) the full vector of all $4^n$ Pauli error rates, denoted as $\chi(\mathcal{E})$, (ii) the vector of length $4^n$ containing $K$ non-zero values, specifically the Pauli error rates from CER, labeled $\chi_{K}(\mathcal{E})$, and (iii) the vector incorporating $K$ Pauli error rates from CER, with the remainder assigned by the \emph{USS} algorithm, represented as $\hat{\chi}_{K}(\mathcal{E})$.

If the distribution in (i) closely resembles (ii), it suggests that the CER data is primarily responsible for the performance enhancement of the ML decoder. To quantify the contribution of the \emph{USS} algorithm, we will estimate the Total Variation Distance (TVD) between the distributions in (i) and (ii), denoted by $\mathrm{TVD}(\chi(\mathcal{E}), \chi_{K}(\mathcal{E}))$. This value will be compared to the TVD between $\chi(\mathcal{E})$ and $\hat{\chi}_{K}(\mathcal{E})$. If the latter is smaller, it indicates that the \emph{USS} algorithm plays a crucial role in enhancing the decoder's understanding of the original noise process beyond the available CER data. As shown in \cref{tab:tvds}, for the ensemble of CG1D error models considered in \cref{fig:cg1d}, $\mathrm{TVD}(\chi(\mathcal{E}), \chi_{K}(\mathcal{E}))$ is approximately five to ten times larger than $\mathrm{TVD}(\chi(\mathcal{E}), \hat{\chi}_{K}(\mathcal{E}))$ for $K = 163$, which represents roughly $1\%$ of the available CER data. This shows that the \emph{USS} algorithm plays a key role in achieving the observed performance gains alongside CER.

\begin{table}
\begin{center}
\begin{tabular}{ |c||c|c|  }
 \hline
 $\langle\epsilon\rangle$ & $\mathrm{TVD}(\chi(\cE), \chi_{K}(\cE)$ & $\mathrm{TVD}(\chi(\cE), \widehat{\chi}_{K}(\cE)$\\
 \hline
 $6.6\times 10^{-3}$ & 0.001 & 0.007   \\
 \hline
 $0.03$ & 0.006 & 0.027 \\
 \hline
 $0.08$ & 0.017 & 0.081\\
 \hline
 $0.19$ & 0.037 & 0.188\\
 \hline
\end{tabular}
\caption{The above table highlights the role of the \emph{Uncorrelated Split Search} (\emph{USS}) algorithm in enhancing the performance of the ML decoder for the ensemble of CG1D error models, as reported in \cref{fig:cg1d}. The second column represents the Total Variation Distance (TVD) between the distribution of Pauli error rates in the noise process, $\chi(\mathcal{E})$, and the distribution of the leading $K = 163$ error rates extracted by CER, denoted as $\chi_K(\mathcal{E})$. In contrast, the last column shows the TVD between $\chi(\mathcal{E})$ and the distribution generated by the \emph{USS} algorithm, $\hat{\chi}_{K}(\mathcal{E})$, which is based on the CER data comprising the leading $K$ Pauli error rates. While the TVD values vary significantly depending on individual instances of the random CG1D error model, we have averaged the results across subsets (bins) of the ensemble, grouped by their process infidelities. The first column represents the average physical error rate for a bin in \cref{fig:cg1d}. The results show that, on average, the \emph{USS} algorithm improves the decoder’s knowledge of the underlying noise process, beyond what is provided by the CER data alone.}
\label{tab:tvds}
\end{center}
\end{table}

\section{Conclusion} \label{sec:conclusion}
In this work, we have demonstrated how the rich set of noise data that can be acquired efficiently using CER can be used directly to improve logical error rates. We demonstrated a method to harness only the $K$ largest Pauli error rates derived from CER to develop an efficient heuristic to reconstruct all of the remaining Pauli error rates. This classical postprocessing method significantly cuts down the experimental cost associated with estimating the error rates for the ML decoder. By integrating our heuristic into the maximum likelihood decoder, we could numerically estimate the suppression in the optimal logical error rate achievable for the given error model. Testing our methods for a wide variety of realistic error models, from coherent errors to incoherent errors, we found a performance gain of at least an order of magnitude when compared to a scenario where only the average error rate of the underlying physical error model is given. The consistent improvement observed across diverse error models underscores the promise of our heuristic approach in completing the Pauli error distribution from limited CER data, regardless of the specific error model chosen.

Although concatenated codes are extensively used in proofs of the fault tolerance accuracy threshold theorem \cite{AGP06}, these are less appealing from a practical standpoint. This is largely attributed to the large weights of the stabilizer generators, resulting in significant errors while measuring an error syndrome. Our focus has been on concatenated codes with increasing distance as levels grow, albeit encoding only a single logical qubit. However, this is not a strict limitation of our methods, and recent works \cite{YTY24,YK24} have shown constructions of concatenated codes that have an increasing rate with the number of levels, and concatenated schemes with topological codes \cite{LKH23} to achieve fault tolerance. Nonetheless, the most promising candidates \cite{Gottesman13,K14,FGL20,KP13,R21} for quantum error correcting codes to achieve fault tolerance with a low resource overhead lie within the realm of quantum Low-Density Parity Check (LDPC) codes, whose stabilizer generators have a weight that is at most a constant independent of the size of the code. Only a handful of quantum LDPC code families have been known to have decoders that can be adapted to the properties of the underlying error model. For instance, Ref.~\cite{STBO18} presents a variant of the standard minimum weight perfect matching algorithm in surface codes \cite{STBO18}, to account for the asymmetry in the probabilities of the single qubit Pauli errors. In some cases, ML decoding algorithms for surface codes based on tensor networks have been adapted to experimental features of the underlying CPTP map \cite{D24,C21}. Analyzing performance gains for adapting a decoder for quantum LDPC codes to limited CER data remains an interesting problem for future research.

The key conceptual implication of our work is that logical error rates can be significantly suppressed when the decoder is provided even a small fraction of the \emph{a priori} error rates. If we consider interpreting these results in the context of FT-QEC, then this would entail accounting for a number of error sources, where memory errors in our analysis serve as a proxy for the \emph{cumulative} errors arising from imperfections in each of the components in fault-tolerant implementations, including the transversal gates on the data qubits, the syndrome extraction circuits across the data and ancilla qubits, the direct PVM measurement of the ancilla qubits, and the recovery operations on the data qubits. While recent work has demonstrated the feasibility and practicality of error learning for particular subsets of the above components of fault-tolerance~\cite{fazio2025characterizingphysicallogicalerrors,hockings2025}, a general (efficient) solution for estimating the cumulative errors under all components required for FT-QEC remains an open problem. 

Our results build a compelling case that the problem of estimating the \emph{cumulative} errors in the fault-tolerant setting presents a key opportunity for optimizing decoder performance. Our work  demonstrates that even a modest amount of the experimentally-accessible error characterization data  can yield significant improvements in decoder performance.

\begin{acknowledgments}
This research was undertaken thanks in part to funding from the Canada First Research Excellence Fund. Research was partially sponsored by the ARO under Grant Number: W911NF-21-1-0007. SDB acknowledges support from the Australian Research Council (ARC) via the Centre of Excellence in Engineered Quantum Systems (EQuS) project number CE170100009. AJ acknowledges support from ERC Starting Grant 101163189 and UKRI Future Leaders Fellowship MR/X023583/1.

\end{acknowledgments}

\begin{appendix}
\section{Cycle Error Reconstruction} \label{sec:app_cer_nr}
In this section, we review the Cycle Error Reconstruction (CER) method \cite{EWPM19,CDHO23} to estimate (i.e.~learn) the Pauli error rates of a CPTP map to multiplicative precision.

We have already encountered the $\chi-$matrix representation of a CPTP map in \cref{eq:chi}. An alternate representation is obtained by expressing the input state $\rho$ as a linear combination of Pauli matrices. The CPTP map $\cE$ can be described as an affine transformation, denoted by $\Lambda(\cE)$, on the vector space of Pauli matrices. The $4^{n}\times 4^{n}$ real matrix $\Lambda(\cE)$ is known as the Pauli Liouville representation \cite{NGKR21}. Each row and column corresponds to a unique pair of $n-$qubit Pauli matrices $P, Q \in \cP_{n}$:
\begin{gather}
\Lambda_{P,Q}(\cE) = \tr(\cE(P)\cdot Q) ~ . \label{eq:ptm}
\end{gather}

Recall the simple yet versatile technique of estimating the average fidelity of a CPTP map using experimental protocols, known as Randomized Benchmarking (RB) \cite{MGE11,MGE12}.
The core concept of the standard RB protocol lies in utilizing the two-design property of the Clifford group \cite{PhysRevA.80.012304}, leading us to rewrite the process fidelity of a CPTP map $\cE$ in \cref{eq:avg_fidelity} as:
\begin{gather}
\epsilon(\cE) = 1 - \dfrac{1}{|\cC_{n}|} ~ \sum_{C \in \cC_{n}}\tr(|0\rangle\langle 0| \cdot C^{\dagger}~\cE(C~\rho~C^{\dagger})~C) ~ , \label{eq:avg_fid_Clifford}
\end{gather}
where $\cC_{n}$ is the $n-$qubit Clifford group. This formulation suggests an experimental design where we initialize a fixed state $|0\rangle$, apply $m$ random Clifford gates $C$ sampled from $\cC_{n}$, followed by an inverse to return to the initial state in the absence of noise, and then measure in the $Z-$basis. The probability of obtaining $+1$, denoted by $P_{s}$, is related to the average fidelity of the underlying CPTP $\cE$ as
\begin{gather}
P_{s} = A_{0}~\epsilon^{m}(\cE) ~, \label{eq:survival_infidelity}
\end{gather}
where $A_{0}$ is a constant related to state preparation and measurement errors.

Note that the quantity measured by RB, $\epsilon(\cE)$, is only one of the $\cO(4^{2n})$ parameters that specify $\cE$. In fact, there are several noise processes of similar fidelity that lead to vastly different logical error rates for quantum error-correcting schemes \cite{IP17}. CER provides a more detailed insight into reconstructing $\cE$ efficiently. In contrast to RB, which provides a single number $r(\cE)$, CER estimates all the $4^{n}$ Pauli error rates in $\cE$.

\begin{figure}
\begin{center}
\includegraphics[scale=0.6]{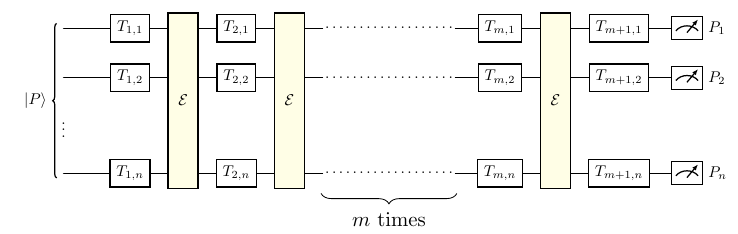}
\caption{The above figure shows a schematic for the CER protocol to estimate a Pauli error rate. The circuit consists of $m$ cycles of the noise $\cE$ followed by random Pauli gates $T_{i} \in \cP_{n}$ expressed as $T_{i} = T_{i,1} \otimes T_{i,2} \otimes \ldots \otimes T_{i,n}$ to perform Twirling. Initializing the input in a Pauli eigen state $|P\rangle$ and measuring at the end in $P$ basis yields a decay law to estimate $\chi_{P,P}(\cE)$.}
\label{fig:cb_schematic}
\end{center}
\end{figure}

Let us define a CPTP map $\tau[\cE]$ obtained by averaging over all possible choices of applying a random Pauli $P$ to both the input and the output of $\cE$:
\begin{gather}
\tau[\cE](\rho) = \dfrac{1}{|\cP_{n}|}~\sum_{P\in\cP_{n}}P~\cE(P~\rho~P)~P ~ . \label{eq:pauli_twirl}
\end{gather}
The map $\tau[\cE]$, called the Pauli Twirl of $\cE$ \cite{ESMR07,MGE11}, is itself a Pauli channel due to a straightforward observation from representation theory. Furthermore, $\cE$ and $ \tau[\cE]$ have identical Pauli error rates. We can now employ a technique akin to RB to extract the Pauli error rates of $\cE$, which is based on the following observation. The effect of $\tau[\cE]$ on an eigenstate of a Pauli operator $P\in\cP_{n}$ with eigenvalue $\lambda_{P}$, denoted by $|P\rangle$, is entirely determined by $\Lambda_{P,P}(\cE)$ defined in \cref{eq:ptm}. Hence, measuring $P$ on $\tau[\cE](|P\rangle\langle P|)$ yields $\lambda_{P}$ with probability that obeys a decay law similar to \cref{eq:survival_infidelity} but with $r(\cE)$ replaced by $\chi_{P,P}(\cE)$. Hence, by choosing the input to be each of the $4^{n}$ Pauli eigenstates, we can estimate all the corresponding Pauli error rates of $\cE$. \Cref{fig:cb_schematic} presents a schematic overview of the CER protocol. It is important to note that \cref{fig:cb_schematic} represents a specific instance of the CER protocol tailored for estimating the Pauli error rates of a memory error process like $\cE$. We will skip detailing the general technique \cite{CDHO23} since it is not directly pertinent to this paper.

The authors in \cite{FW20} show that all of the Pauli error rates can be estimated using $\cO(n 2^{n})$ measurements in total and employing the Walsh-Hadamard transformation to convert the eigenvalues to the Pauli error rates. While CER offers a scalable and efficient technique to extract each of the $4^{n}$ Pauli error rates for a $n-$qubit CPTP map, the exponential growth in their number makes it impractical to estimate all of them within polynomial time. This leads us to another development, presented in \cite{HYF21} where the authors propose a method to estimate the $K-$largest Pauli error rates of a CPTP map $\cE$, using a number of experiments that scale linearly with $K$. Our work leverages the result of \cite{HYF21}.

\section{Quantum Error Correction Theory for Non-Pauli Noise} \label{sec:qec_non_Pauli}
Recall that a $[[n,k,d]]$ stabilizer code \cite{GotPhD97} $\cQ$ is defined as a $2^{k}$ dimensional subspace of $\bC^{n}_{2}$, such that every element of $\cQ$ is an eigenstate with eigenvalue $+1$ of $n-k$ commuting Pauli operators called the stabilizer generators.

Before describing the QEC strategy, let us first introduce the necessary notation. Let $\rho$ denote a single qubit state which is encoded into a $[[n,k,d]]$ stabilizer code whose encoding circuit \cite{MP24} is $U$, yielding  $\ol{\rho}$ given by:
\begin{gather}
\ol{\rho} = U \cdot \rho \otimes |0\rangle\langle 0|^{n-1} \cdot U^{\dagger} ~ . \label{eq:encoded_state}
\end{gather}
A noisy encoded state is modeled by the action of a $n-$qubit CPTP map $\cE$ on $\ol{\rho}$ resulting in $\cE(\ol{\rho})$.

The strategy for dealing with general noise is inspired by the Pauli case. In the simple case where $\cE$ is a Pauli error model described in \cref{eq:pauli_channel}, we note that any Pauli error $E$ on the encoded state may be detected by observing the outcomes of measuring each of the stabilizer generators, or equivalently, by examining the commutation relations between $E$ and each of $\{g_{1}, \ldots, g_{n-k}\}$. The binary sequence $s \in \bZ^{n-k}_{2}$ where $s_{i} = -1$ when $[E, g_{i}] = 0$ and $-1$ when $\{E, g_{i}\} = 0$, is called \emph{error syndrome}. Clearly, there are only $2^{n-k}$ distinct error syndromes, whereas $4^{n}$ distinct $n-$qubit Pauli errors. So, many Pauli errors will have the same error syndrome. The Pauli channel now induces a probability distribution on the error syndromes where the probability of an error syndrome $s$, denoted by $\Prob(s)$ is given by the sum of probabilities of all Pauli errors that result in the error syndrome $s$.

The case of non-Pauli errors is somewhat more complicated. Measuring each stabilizer generator $g_{i}$ on $\cE(\rho)$ yields an error syndrome $s$ with probability $\Prob(s)$ given by
\begin{gather}
\Prob(s) = \tr(\Pi_{s} \cdot \cE(\ol{\rho})) ~ , \label{eq:prob_s}
\end{gather}
where $\rho$ is the encoded state and $\Pi_{s}$ is a projector corresponding to the syndrome measurement:
\begin{gather}
\Pi_{s} = \prod_{i = 1}^{n-k}\dfrac{\bI + (-1)^{s_{i}}}{2} ~ . \label{eq:proj_s}
\end{gather}
Expressing the noisy state $\cE(\ol{\rho})$ using the chi-matrix in \cref{eq:chi}, we find that
\begin{gather}
\Prob(s) = \sum_{P,Q \in \cP_{s}}\chi_{P,Q}(\cE) \tr(\Pi_{s} \cdot P \cdot \cE(\ol{\rho})\cdot Q) ~ , \label{eq:prob_s_chi}
\end{gather}
where $\cP_{s}$ is the set of all Pauli errors whose error syndrome, obtained from commutation relations with stabilizer generators, is $s$. Contrary to a Pauli error model, the probability of an error syndrome in non-Pauli error models is not independent of the input state. Additionally, computing $\Prob(s)$ from \cref{eq:prob_s_chi} requires time proportional to $2^n$, unlike the simple case of Pauli channels, where it only takes time quadratic in $n$. After measuring an error syndrome, the state of the system transforms to
\begin{gather}
\dfrac{\Pi_{s} \cdot \cE(\ol{\rho}) \cdot \Pi_{s}}{\Prob(s)} ~ . \label{eq:post_measurement_s}
\end{gather}

\subsection{Maximum Likelihood Decoding} \label{sec:ml_decoding}
Before defining the decoding problem, it is worth analyzing some structure of the Pauli group induced by the stabilizer group. Note that there are non-trivial Pauli errors $E \neq \bI$ which have a trivial syndrome: $s = 00\ldots 0$; these errors belong to $\cN(\cS)$: the normalizer of the stabilizer group $\cS$ within the Pauli group $\cP_{n}$.
\begin{gather*}
\cN(\cS) = \{E ~ \in~ \cP_{n} ~ : ~ E \cdot S\cdot E ~\in~ \cS\} ~ .
\end{gather*}
Clearly, $\cS\subset \cN(\cS)$, which encompasses errors that leave all encoded states unchanged. However, errors in $\cL = \cN(\cS) / \cS$, referred to as logical errors, are those causing non-trivial transformations to the encoded states yet cannot be detected by the error syndrome.  Finally, $\cT = \cP_{n} / \cN(\cS)$, referred to as pure errors \cite{PP13,IP15,Beale23}, denotes classes of errors where each class corresponds to a unique error syndrome. Consequently, we can now express any Pauli error $E \in \cP_{n}$ as a product: $E = T \cdot L \cdot S$ where $T \in \cT, L \in \cL$, and $S \in \cS$.

In the decomposition of a Pauli error shown in \cref{eq:tls}, note that each pure error generator anticommutes with a unique stabilizer generator. In other words, the error syndrome tells us precisely the pure error generators that make up the $T-$element of the error:
\begin{gather}
T = \prod_{i=1}^{n-k}T^{s_{i}}_{i} ~ . \label{eq:Ts}
\end{gather}
While $T$ is entirely fixed by learning the error syndrome $s$ through measurements: $T \equiv T_{s}$, it remains to determine $L$ and $S$ to completely specify a Pauli error. The choice of $S$ is not important, as two errors that differ in this decomposition by a stabilizer are effectively indistinguishable. However, the choice of the logical operator $L\in\cL$ is crucial--applying a wrong choice will not return us to the right encoded state after QEC, marking its failure.

The task of determining $L$ is a statistical inference problem known as Maximum Likelihood (ML) decoding \cite{HL11,IP15}, stated as follows. Given an error syndrome $s$ and an error model $\cE$, compute $L_{s} \in \cL$ such that
\begin{gather}
L_{s} = \argmax{L \in \cL}~\Prob(L | s) ~ , \label{eq:MLD_prob_L}
\end{gather}
where $\Prob(L | s)$ is the probability of a logical error, defined as the sum of all errors in the coset $T_{s}\cdot L\cdot \cS$:
\begin{gather}
\Prob(L | s) = \sum_{S \in \cS}\Prob(T_{s}\cdot L\cdot S) ~ . \label{eq:prob_Ls_app}
\end{gather}
Note that the probabilities of Pauli errors $P = T_{s}\cdot L\cdot S$ are derived from a given error model. If $\cE$ is a Pauli error model, this is straightforward $\Prob(P) = \chi_{P,P}(\cE_{\cP})$. In the general case, including non-Pauli error models, we derive $\Prob(P)$ from the Pauli Twirl of $\cE$: $\tau[\cE]$. So, $\Prob(P) = \chi_{P,P}(\tau[\cE])$.

An alternative decoding strategy involves selecting a logical operator that maximizes the probability of a specific error for a given choice of $S$, i.e.
\begin{gather*}
L_{s} = \argmax{L~\in~\cL,~ S~\in~\cS} ~ \Prob(T_{s} ~L ~ S) ~.
\end{gather*}
For an i.i.d. depolarizing noise model with an error rate $p$ per qubit, the operation $T_{s}~L_{s}~ S$ represents the lowest weight Pauli error that is consistent with the observed syndrome $\vec{s}$. This approach defines \emph{Minimum Weight (MW) Decoder} \cite{IP15}.

The output of the decoder: $L_{s}$, can be combined with $T_{s}$ and any stabilizer $S$ to yield a recovery operation: $T_{s} \cdot L_{s}\cdot S$. After applying the recovery operation on the $n-$qubit state in \cref{eq:post_measurement_s}, we find
\begin{gather}
\ol{\sigma} = \dfrac{\left(T_{s} \cdot L_{s} \cdot S \cdot \Pi_{s} \cdot \cE(\ol{\rho}) \cdot \Pi_{s}\cdot S \cdot L_{s} \cdot T_{s}\right)}{\Prob(s)} ~ . \label{eq:post_recovery}
\end{gather}
Since syndrome measurement and recovery involve active quantum controls, errors are inevitable during these steps. However, we will restrict ourselves to a setting in which noise in the quantum error correction circuit is significantly suppressed compared to noise in quantum memory \cite{KMF24,COT07,MCG17,Cross08,CWBL17}. In such a case, the state in \cref{eq:post_recovery} is guaranteed to be an encoded state. Consequently, we can extract the encoded logical information, denoted by $\sigma$, from $\ol{\sigma}$:
\begin{gather}
\sigma = \tr_{2, \ldots n}\left(U^{\dagger}\cdot \ol{\sigma} \cdot U\right) ~ , \label{eq:unencoding}
\end{gather}
where $\tr_{2, \ldots, n}$ represents a partial trace over the $2^{n-1}-$dimensional subspace of all qubits except the first one. The convenience of assuming only memory errors enables us to define a CPTP map $\cE^{s}_{1}$ that maps the logical information $\rho$ to $\sigma$. This map is called \emph{effective channel} \cite{RDM02,CWBL17,GB15,IP17,DP17,DP18}, as it describes the effective action of noise along with the correction of quantum errors on the logical information. The average effective channel, denoted by $\ol{\cE}_{1}$ is simply the average of $\cE^{s}_{1}$ overall error syndromes $s$:
\begin{gather}
\ol{\cE}_{1} = \sum_{s}~\Prob(s)~\cE^{s}_{1} ~ . \label{eq:avg_eff_chan}
\end{gather}
We will refer to the strength of noise in the logical channel, in short, \emph{logical error rate}, as the infidelity of the average effective channel \cite{IJBE22}: $\epsilon(\ol{\cE}_{1})$, in short, $\ol{\epsilon}_{1}$.

\subsection{Efficient Decoding of Concatenated Codes} \label{sec:decoding_concatenated}
While ML decoding is optimal \cite{HL11,IP15} for stabilizer codes, it comes with certain trade-offs. Firstly, it necessitates the knowledge of all the $4^{n}$ Pauli error rates affecting the physical qubits of the code. Despite this practical roadblock, we will analyze the performance of ML decoding since it yields the best logical error rate achievable under a given physical error model. Secondly, ML decoding poses computational challenges for general stabilizer codes \cite{IP15}. However, to circumvent this, we narrow the focus to concatenated quantum codes, a rare family of stabilizer codes that have an efficient ML decoder implemented using a \emph{message passing} algorithm \cite{P06}.

\begin{figure}
\begin{center}
\includegraphics[scale=0.4]{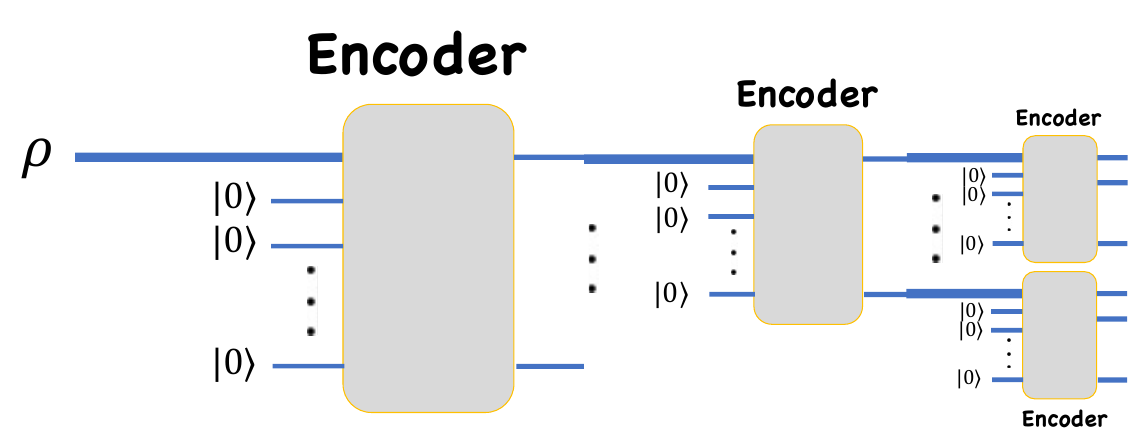}
\caption{Figure showing the encoding circuit of a concatenated code. Each code block is represented by an encoder, which encodes one logical qubit into $n$ physical qubits. Note that the encoding circuit resembles a tree. Each layer of encoder operations is called a concatenation level, and the code shown above is a level$-3$ concatenated code.}
\label{fig:schematic_encoder_concatenated}
\end{center}
\end{figure}

Concatenated codes employ recursive encoding strategies in which the encoding circuit resembles a tree \cite{GotPhD97,YTY24,YK24}. \Cref{fig:schematic_encoder_concatenated} shows a schematic of the circuit circuit. Each block in the $\ell$ layers denotes a $[[n,1,d]]$ stabilizer code, resulting in a $[[n^{\ell}, 1, d^{\ell}]]$ stabilizer code. The code is referred to as a level$-\ell$ concatenated code. The error syndrome for any Pauli error $E$ comprises $n^{\ell}-1$ bits, which can be grouped into sequences corresponding to each code block. Assuming that the noise $\cE$ is uncorrelated across qubits in different code blocks, the probability of a logical operator denoted as $\Prob(L | s)$ in \cref{eq:prob_Ls_app}, takes on a product form \cite{P06}.

\begin{figure}
\begin{center}
\includegraphics[scale=0.05]{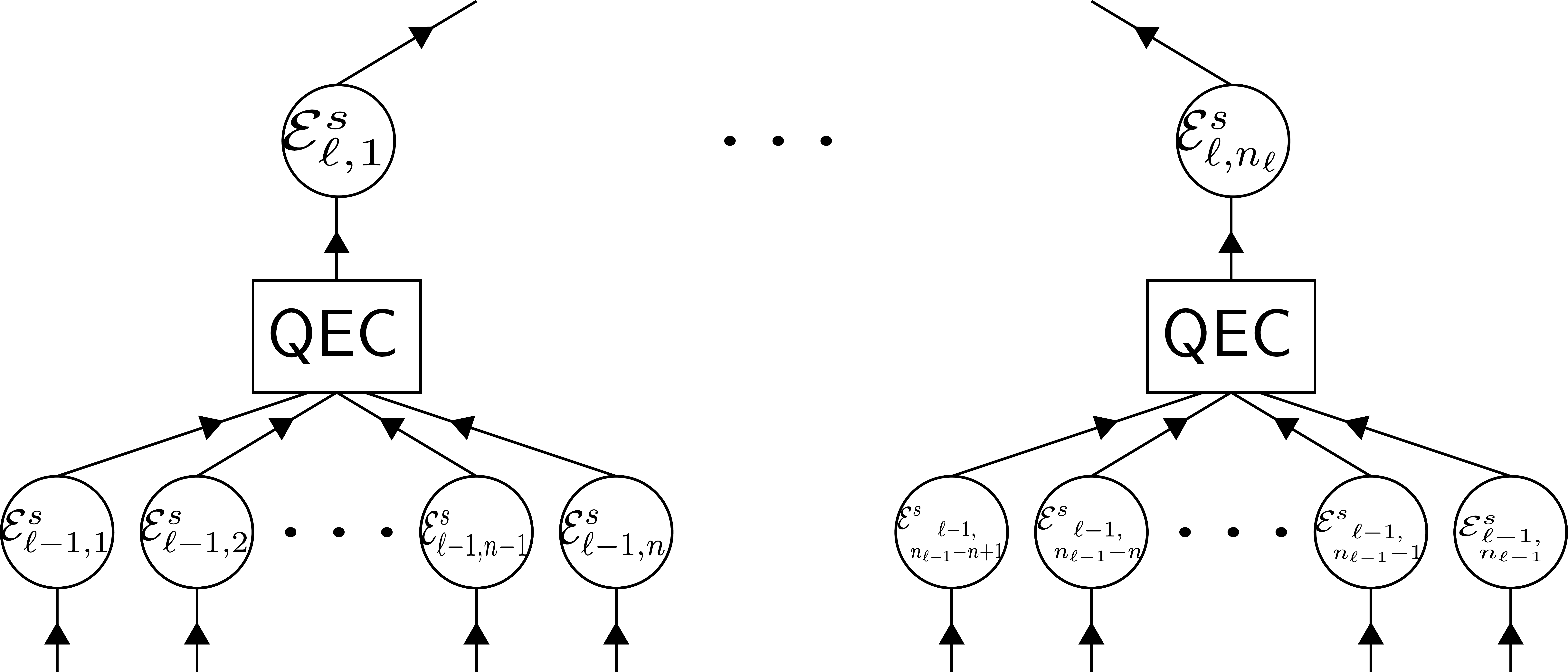}
\caption{Figure showing the operation of the message passing decoder for concatenated codes. The QEC block for a code block at level $\ell$ encapsulates all the steps to compute an effective logical channel conditioned on the error syndrome of the corresponding code block, using as the physical noise processes the logical channels computed from code blocks at level $\ell-1$.}
\label{fig:schematic_message_passing}
\end{center}
\end{figure}

Consequently, the ML decoder operates as follows. To begin with, we use \cref{eq:prob_Ls_app} to compute the distribution of logical operations on each of the $n^{\ell-1}$ code blocks at level$-1$, conditioned on the syndrome bits of the code block, independently. Hence, for each codeblock we obtain a distribution $\{\Prob(\ol{I} | s), \Prob(\ol{X} | s), \Prob(\ol{Y} | s), \Prob(\ol{Z} | s)\}$, and a logical channel $\cE^{s}_{1}$. Each of these logical channels serves as the physical noise channel to the compute logical channel for the code blocks at level$-2$. Hence, we compute the logical channels for all the code blocks at level$-2$, conditioned on the syndrome bits of the code blocks, independently. We now have to compute a distribution for the logical corrections for each of the code blocks at the level$-2$. For this, we use \cref{eq:prob_Ls_app}, where the Pauli error model to be used to derive the probabilities of physical $I, X, Y, Z$ errors is replaced by the distribution of logical errors from the level$-1$ code blocks. We proceed in this iterative fashion until the final level, where the distribution of logical operators for the code block at level$-\ell$ is computed. Ultimately, the output of the ML decoder is the logical operator with the highest probability, described in \cref{eq:MLD_prob_L}. Applying this logical operator allows us to compute an effective channel for the level$-\ell$ concatenated code using \cref{eq:post_recovery,eq:unencoding,eq:avg_eff_chan}. \Cref{fig:schematic_message_passing} provides a schematic for the functioning of the ML decoder for concatenated codes.

For a concatenated code with $\ell$ levels, we can follow the recipe described above to compute the effective logical channel conditioned on the observed syndrome, denoted $\cE^{s}_{\ell}$. The average effective channel, $\ol{\cE}_{\ell}$, which we refer to in short as $\ol{\epsilon}_{\ell}$, is obtained by averaging over all possible syndrome outcomes, similar to \cref{eq:avg_eff_chan}. The average logical error rate is then defined as the infidelity of $\ol{\cE}_{\ell}$. However, the number of syndrome outcomes grows exponentially with the number of levels $\ell$, making an exact computation of the average effective channel impractical. For instance, even at $\ell=2$, there are approximately $10^{12}$ possible syndrome outcomes. \Cref{sec:numerics} of the appendix details efficient sampling techniques for estimating the average logical error rate for concatenated codes, bypassing the need to explicitly enumerate all syndrome outcomes.

\section{Random Unitary Circuits} \label{sec:random_unitary}
Coherent errors constitute an important class of physically relevant error models, arising from various scenarios such as imperfect control during logical operations or unintended qubit interactions within a code. These are particularly malicious for any quantum code due to two key characteristics. Firstly, they accumulate more rapidly than incoherent errors. Coherent errors, being rotations, add linearly, unlike incoherent errors, which only accumulate quadratically \cite{HDF19,GD17}. Secondly, since most quantum codes are tailored to correct incoherent errors such as depolarizing noise, their effectiveness decreases when faced with coherent errors \cite{BEKP18}. For most of the realistic hardware, it has been observed that errors are often correlated in spatial as well as temporal regions. With a loss of generality, we want to focus on time-independent yet spatially correlated unitary errors. For instance, when we apply a two-qubit entangling gate such as CNOT between two qubits in a code, we observe that a faulty implementation leaves behind correlated errors \cite{ZLLX22,TCKY22}. In superconducting architectures, cross-talk is a major source of noise, which is often of similar magnitude as the single qubit errors \cite{KW23}. Since unitary errors take a specific form depending upon the choice of hardware device under consideration, we will analyze random correlated unitary error models so that our outstanding conclusions apply to any realistic hardware.

In this section, we showcase gains for unitary errors resulting from a random unitary circuit of a fixed depth \cite{CBQA20,LLAJ22}. Recall that the gain refers to a relative performance gain, defined in \cref{eq:def_delta}, of the decoder with access to $K$ error rates in the CER data, over a decoder that assumes access to the average physical error rate alone.

We consider a model for coherent errors wherein an $n$-qubit unitary error $U$ is composed of at most $d$ random unitary errors $U_{1}, \ldots, U_{d}$, each acting non-trivially on at most $k \leq n$ qubits. For convenience, let us label the support of $U_{i}$ by the set $\mathfrak{S}_{i}\subseteq \{1, \ldots, n\}$. We model each of the unitary errors $U_{i}$ by first choosing $\mathfrak{S}_{i}$ and then sampling a random unitary matrix of dimension $2^{\left|\mathfrak{S}_{i}\right|}$. Typically, a physical system has a mean correlation length $\lambda$, indicating that most errors in ${U_{1}, \ldots, U_{d}}$ are supported on $\lambda$ qubits. Additionally, we expect that the frequency of $k-$qubit errors to decrease exponentially with $k$. These assumptions are reflected in our model by sampling $\{\left|\mathfrak{S}_{1}\right|, \left|\mathfrak{S}_{2}\right|, \ldots, \left|\mathfrak{S}_{d}\right|\}$ from a Poisson distribution with mean $\lambda$. Once $\left|\mathfrak{S}_{i}\right|$ is chosen, we assign $\mathfrak{S}_{i}$ to a random subset of $\{1, \ldots, n\}$ of the given size. Finally the random unitary matrix $U_{i}$ is obtained by exponentiating a random Hermitian matrix $H_{i}$ of dimension $2^{\left|\mathfrak{S}_{i}\right|}$ in the form: $U_{i} = \exp\left(i ~ H_{i} ~ t\right)$, where $t$ is a virtual time parameter which is fixed to acquire a handle on the strength of noise in the system. To introduce significant correlations in our system, we set $\lambda = 2$ for generating random unitary errors in this model. Note that for these correlated error models, a minimum weight decoder is likely to fail since multi-qubit errors are not exponentially suppressed compared to single-qubit errors. Furthermore, we expect that learning the actual probabilities of the multi-qubit errors from the limited CER data would be crucial since they would deviate significantly from our ansatz for the multi-qubit error probabilities assigned using the \emph{USS} method in \cref{alg:uss}.

\begin{figure*}
\begin{center}
\includegraphics[width=\linewidth]{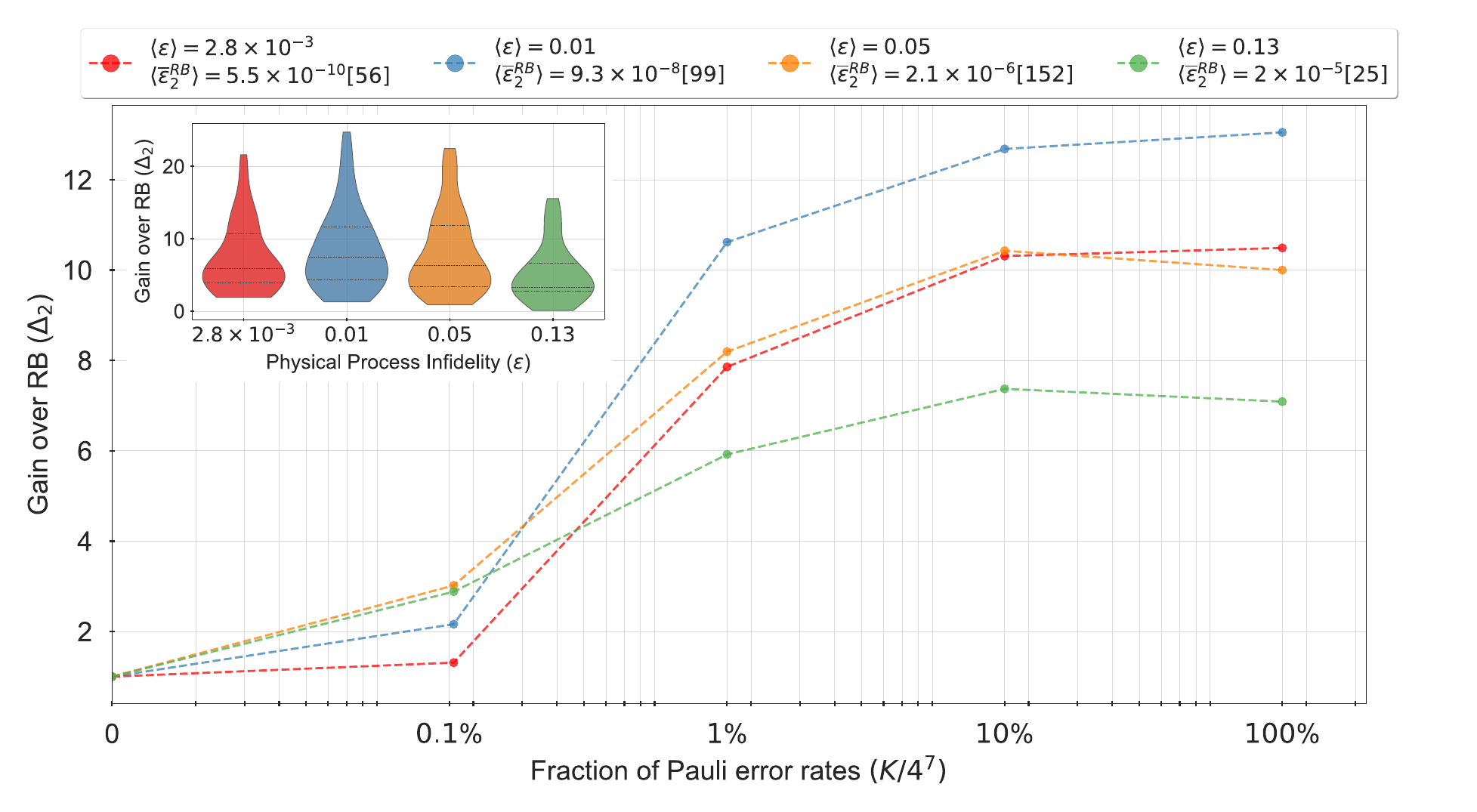}
\end{center}
\caption{The above figure illustrates the performance improvements achieved by the ML decoder for concatenated Steane codes, quantified by $\Delta_{2} = \ol{r}_{2}(\cE ~ ; ~ \cD_{1})/\ol{r}_{2}(\cE ~ ; ~ \cD_{K})$, under unitary errors, with varying fractions of CER data provided to the decoder. Each unitary error is generated from a random unitary circuit of depth at most $9$, consisting of random single, two, and three qubit gates. The error maps are grouped into bins as described in \cref{fig:cg1d}. A typical performance gain of $5X$ can be achieved in most noise regimes, leveraging merely 1\% of the error characterization data from CER, i.e., 164 Pauli error rates. The inset plot shows the variation of performance gains for the individual channels in each of the bins, when only 1\% of the error data is available to the ML decoder. While the median performance gains in the bins are $5X$, we see that for some unitary errors, the performance gains can be as high as $20X$.}
\label{fig:unitary}
\end{figure*}

\Cref{fig:unitary} shows gains for the level$-2$ concatenated Steane code achieved using an ML decoder that leverages Pauli error rates in $\cN(K)$ along with the \emph{USS} method for several values of $K$. The trends in the figure are drawn from an ensemble of $336$ random unitary errors. To generate each unitary error, we build a random unitary circuit of depth $12$ by choosing random unitary gates supported on random subsets of the qubits in the Steane code block whose sizes are distributed according to the Poisson distribution with mean two. Recall that each unitary gate is obtained by exponentiating a random Hamiltonian multiplied by a constant $t$, providing a handle on the noise strength. To generate the random channels shown in \cref{fig:unitary} we have considered six distinct values of $t$ the range $[0.25, 0.65]$. These numbers are chosen to obtain the desired spectrum in the fidelities of the resulting $n-$qubit unitary errors.

\section{Generic Markovian Errors} \label{sec:cptp_error_model}
A slightly inclusive notion of noise processes that includes decoherence effects is given by CPTP maps. An $n-$qubit CPTP map can be described using at most $\cO(2^{n^{2}})$ operators:
\begin{gather}
\cE(\rho) = \sum_{i}K_{i} ~\rho ~ K^{\dagger}_{i} ~ , \label{eq:kraus_decomp}
\end{gather}
where $K_{i}$ are positive matrices, called \emph{Kraus operators} \cite{L19} and satisfy
\begin{gather*}
\sum_{i}K^{\dagger}_{i}~K_{i} = \bI ~ .
\end{gather*}
The $\chi-$matrix representation of $\cE$ shown in \cref{eq:chi} can be derived by expressing the Kraus operators in \cref{eq:kraus_decomp} as linear combinations of Pauli matrices.

An important subclass of CPTP maps are \emph{non-catastrophic channels}, which are characterized by $\epsilon(\cE) \geq 1/2$ and $u(\cE) \geq 1/2$, where $u(\cE)$ is the unitarity \cite{WGHF15} of the CPTP map $\cE$. In this study, we focus specifically on non-catastrophic CPTP maps, which we refer to as realistic errors. It can be shown that a non-catastrophic channel $\cE$ can be accurately described using only one Kraus operator \cite{CAE19}: $\cE(\rho) \simeq K~\rho~ K^{\dagger}$, where $K$ is the \emph{leading Kraus operator}. Expressing $K$ in terms of its polar decomposition yields $K = P ~ U$, where $U$ is a unitary error while $P$ is a projector, denoting an incoherent error. This decomposition highlights a crucial point: realistic CPTP maps can be effectively described as a combination of a unitary error and an incoherent error. We will exploit this idea to sample realistic error models by composing a unitary error $\cE_{\cU}$ in \cref{eq:unitary_channel} with a Pauli channel $\cE_{\cP}$ in \cref{eq:pauli_channel}.

We construct random $n-$qubit CPTP maps by composing at most $m$ CPTP maps on $k \leq n$ qubits. To derive one such CPTP map, first sample a number $u$ from a Poisson distribution with mean $\lambda$, indicating a mean correlation length in the system. Then we compute a random subset of $n$ qubits of size $u$, denoted by $\mathfrak{S}_{i}$, and generate two maps supported on $\mathfrak{S}_{i}$. One is a random Unitary error expressed as $\cE_{\cU}(\rho) = \exp(i~H~t)~\rho~\exp(-i~H~t)$ where $H$ is a random Hermitian matrix of dimension $2^{u}$. The other is a random Pauli channel $\cE_{\cP}$ of fixed average physical error rate, obtained by choosing a random vector of $4^{n}-1$ real numbers denoting the probabilities of all the nontrivial Pauli errors supported on $\mathfrak{S}$. Finally, we compose $\cE_{\cU}$ and $\cE_{\cP}$ to obtain one instance of a random non-catastrophic CPTP map. Composing several of these random CPTP maps yields an $n-$qubit CPTP map.

\begin{figure*}
\begin{center}
\includegraphics[width=\linewidth]{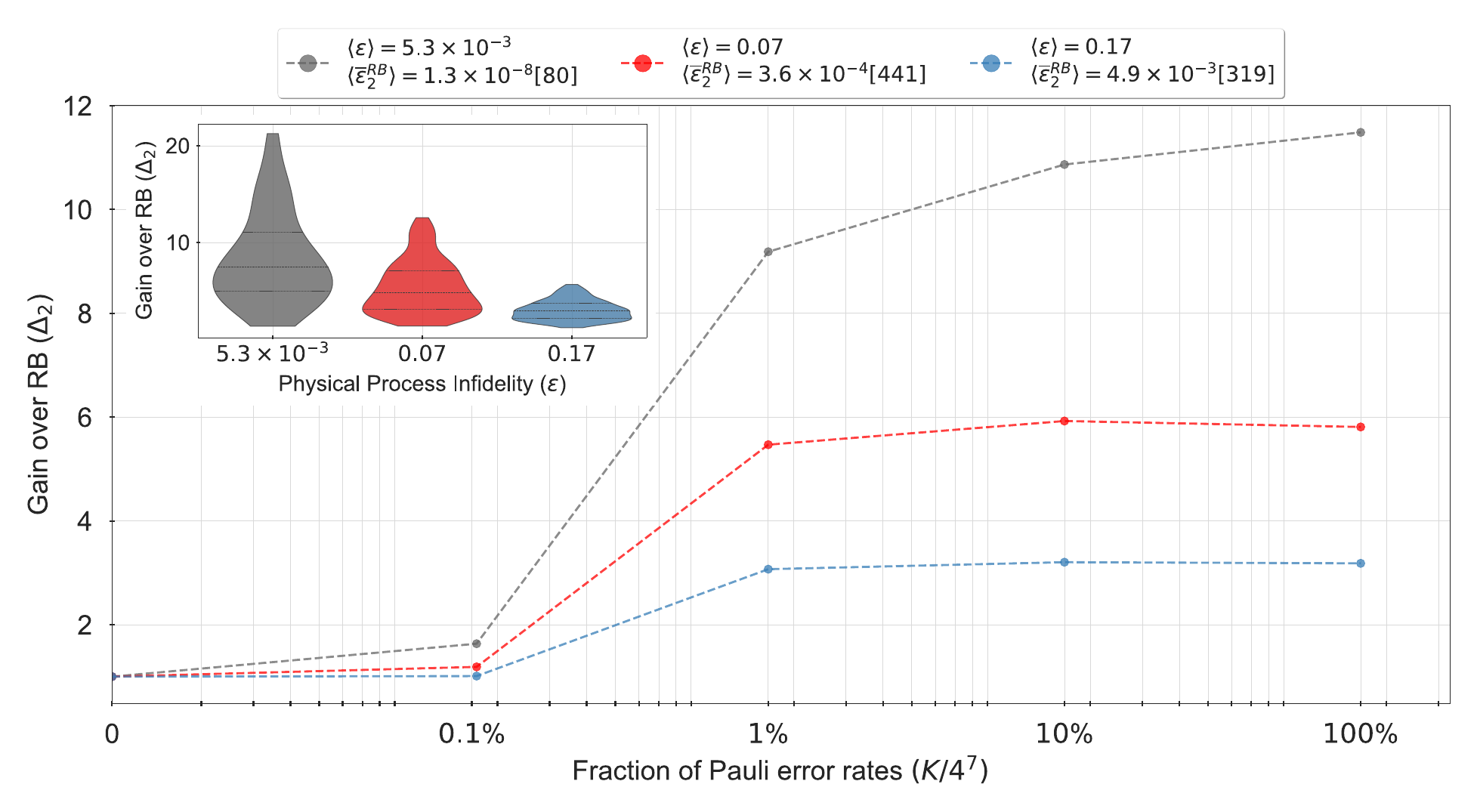}
\end{center}
\caption{The above figure illustrates the performance improvements achieved by the ML decoder for concatenated Steane codes, quantified by $\Delta_{2} = \ol{r}_{2}(\cE ~ ; ~ \cD_{1})/\ol{r}_{2}(\cE ~ ; ~ \cD_{K})$, under random CPTP noise processes, with varying fractions of CER data provided to the decoder. These maps are grouped into bins using a similar strategy as described in \cref{fig:cg1d,fig:unitary}. Across most noise regimes, a $5X$ to $10X$ performance gain is observed, even when only 1\% of the CER data is made available to the ML decoder. Each violin-shaped region in the inset corresponds to a bin identified by its respective color. The width of the region at a specific height ($\Delta_{2}$) is proportional to the number of random CPTP maps that exhibited the corresponding performance gain, when only $1\%$ of the error characterization data from CER is made available to the ML decoder. While the median performance gain across channels in a bin typically ranges between $5X$ and $10X$, individual maps exhibit performance gains as high as $20X$.}
\label{fig:cptp}
\end{figure*}

Similar to the analysis of unitary errors, we examine typical performance gains for random CPTP maps categorized into bins based on their physical infidelities. \Cref{fig:cptp} shows $\approx 10X$ performance gains achieved for the level$-2$ concatenated Steane code across several ensembles of non-catastrophic CPTP maps, using only $1\%$ of the Pauli error probabilities in the CER data.

Together with the unitary cases shown in \cref{fig:cg1d,fig:unitary}, the results in \cref{fig:cptp} underscore the significance of CER data and demonstrate the versatility of the Uncorrelated Split Search method in achieving substantial performance gains.

\section{\texorpdfstring{$k$}{}-body Noise reconstruction}
\begin{figure*}
\begin{center}
\subfloat[\label{subfig:a}]{%
  \includegraphics[width=0.9\linewidth]{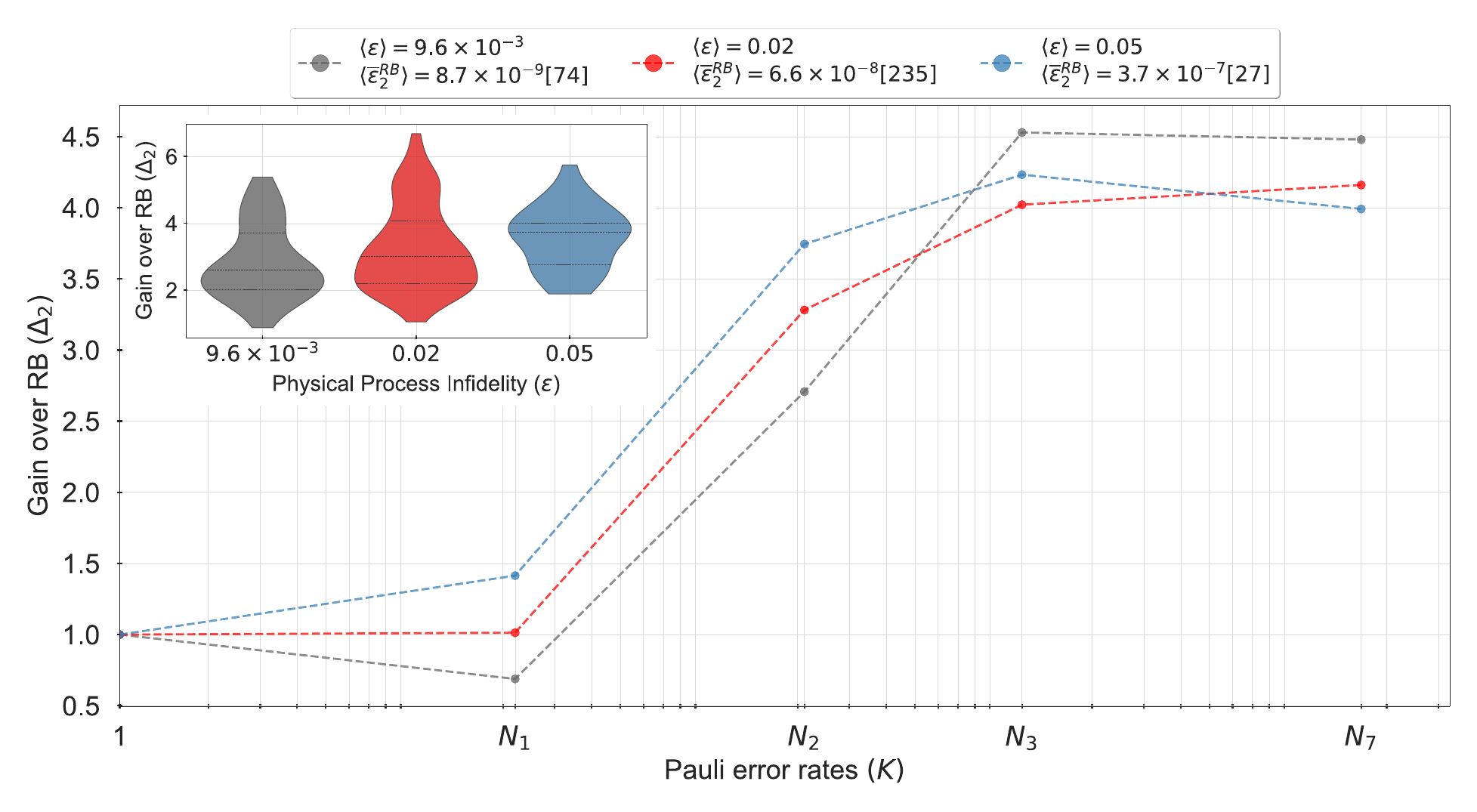}%
}

\subfloat[\label{subfig:b}]{%
  \includegraphics[width=0.9\linewidth]{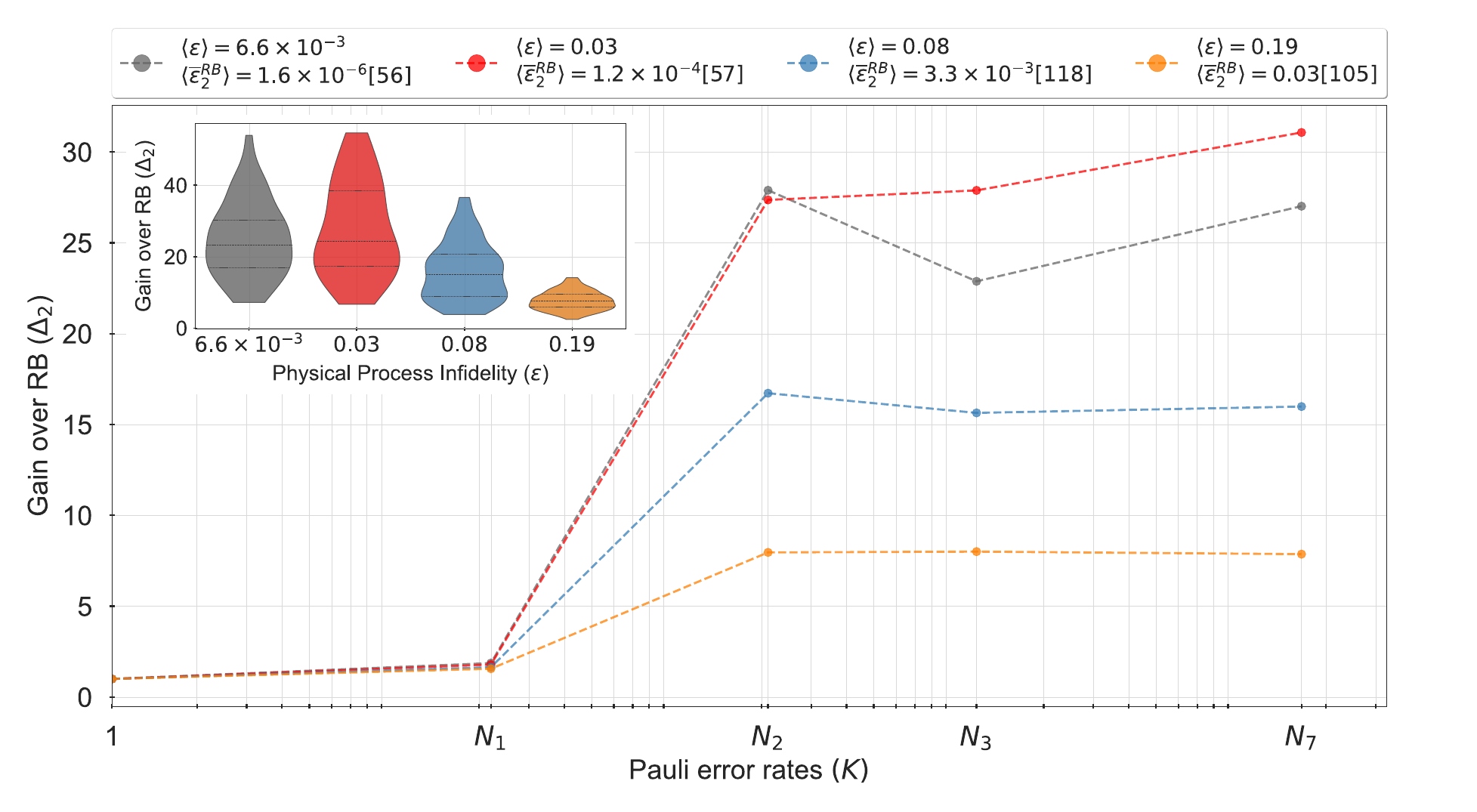}%
}

\caption{The figure above highlights the significance of calibration data from $k$-body Noise Reconstruction ($k-$NR) in enhancing the performance of an ML decoder for concatenated Steane codes under two distinct models for coherent errors. In the top plot, unitary errors are generated from random unitary circuits with depths up to 9, comprising random single- and two-qubit gates. In contrast, the bottom plot uses the CG1D error model to generate the errors. The $X$-axis represents the number of Pauli error rates ($K$) accessible to the decoder, specifically all the Pauli error rates for Pauli errors of weight up to $w$ for $0\leq w\leq 3$ denoted by $N_w$. Performance improvements are quantified using $\Delta_{2} = \ol{r}_{2}(\cE ~ ; ~ \cD_{1})/\ol{r}_{2}(\cE ~ ; ~ \cD_{K})$. In the top plot, for each bin containing random noise processes with similar average physical error rates, the median performance gain is observed to be approximately $5X$. Meanwhile, for the CG1D error model in the bottom plot, the median gains are significantly larger, typically around $30X$. As shown in \Cref{fig:cg1d,fig:unitary}, the inset plots detail the individual performance gains for all noise processes across the bins when the ML decoder is provided with all single- and two-qubit error rates. Notably, these individual gains can far exceed the median values. Overall, where calibration data is sourced from $k-$NR, the performance gains are not as pronounced compared to when the calibration data comprises the $K$ largest Pauli error rates. This disparity arises from the prevalence of correlated errors in the physical noise model, which causes some low-weight errors to occur less frequently than certain higher-weight errors.}
\label{fig:knr_gains}
\end{center}
\end{figure*}

Recall that the goal of this paper is to highlight the critical importance of noise calibration, even when it constitutes as little as 1\% of the total available spectrum, for substantially enhancing the performance of the ML decoder. Our numerical findings in \cref{sec:results} are derived based on a particular selection of limited error characterization information from CER: representing the $K-$largest Pauli error rates in the underlying physical noise process, which can be obtained using experimental protocols detailed in \cite{FW20,HF23}. However, the scope of our techniques for assigning the rates for the remaining errors, using the \emph{Uncorrelated Split Search} algorithm, is agnostic to the choice for the limited CER information. In this section, we stress this point employing a different rationale for selecting partial CER data based on the Hamming weight of Pauli errors, known as $k-$body Noise Reconstruction, or $k-$NR as referenced in \cite{EWPM19,CDHO23}. Here, we describe our partial calibration data for an $n-$qubit CPTP map as comprising Pauli error rates of all Pauli errors with a weight of at most $k$, denoted by $\chi_{k}(\cE)$. Note that the size of $\chi_{w}(\cE)$ equals $N_w$ as shown in \cref{fig:knr_gains}, which scales exponentially with $w$. For example, $k=3$ denotes calibration data consisting of $N_3$ Pauli error rates, corresponding to single-qubit, two-qubit and three-qubit Pauli errors. \Cref{fig:knr_gains} illustrate the performance gains achieved with increasing amounts of $k-$NR data, where the remaining Pauli error probabilities are assigned using the \emph{Uncorrelated Split Search} algorithm. Notably, we observe $\approx 5X$ performance improvement for unitary errors when the decoder is provided with single- and two-qubit error rates, and $\approx 10X$ improvement for CG1D errors. To have a fair comparison of the two different rationales for selecting the limited CER data given to the ML decoder, we utilized the same ensemble of physical noise processes in \cref{fig:knr_gains} as in the numerical studies for \cref{fig:cg1d,fig:unitary}.

In most physical models, it is typically expected that interactions involving a significant number of qubits tend to be weaker than those involving fewer qubits. Essentially, this implies that the likelihood of Pauli errors decreases exponentially with error weight once it exceeds a certain correlation length $\lambda$ within the system. This rule of thumb supports using $k-$NR data for decoding purposes. However, when $k \leq \lambda$, it is possible that some Pauli errors of higher weight are more probable than certain errors of lower weight. Another case where the effective correlation length might increase is when noise accumulates over time, i.e. under composition. In such scenarios, it is more pertinent to focus on the largest error rates than to choose systematically by weight. This is evident from the larger performance improvements observed in the composition of the unitaries in \cref{fig:unitary} versus the top plot in \cref{fig:knr_gains}. In contrast, the absence of inherent composition in the error model leads to the performance gains from the $k-$NR data depicted in the bottom plot of \cref{fig:knr_gains} being similar to those obtained by choosing the leading Pauli error rates for the same physical noise processes within the CG1D error model.

\section{Estimating the average logical error}
\begin{figure*}
\begin{center}
\includegraphics[scale=0.15]{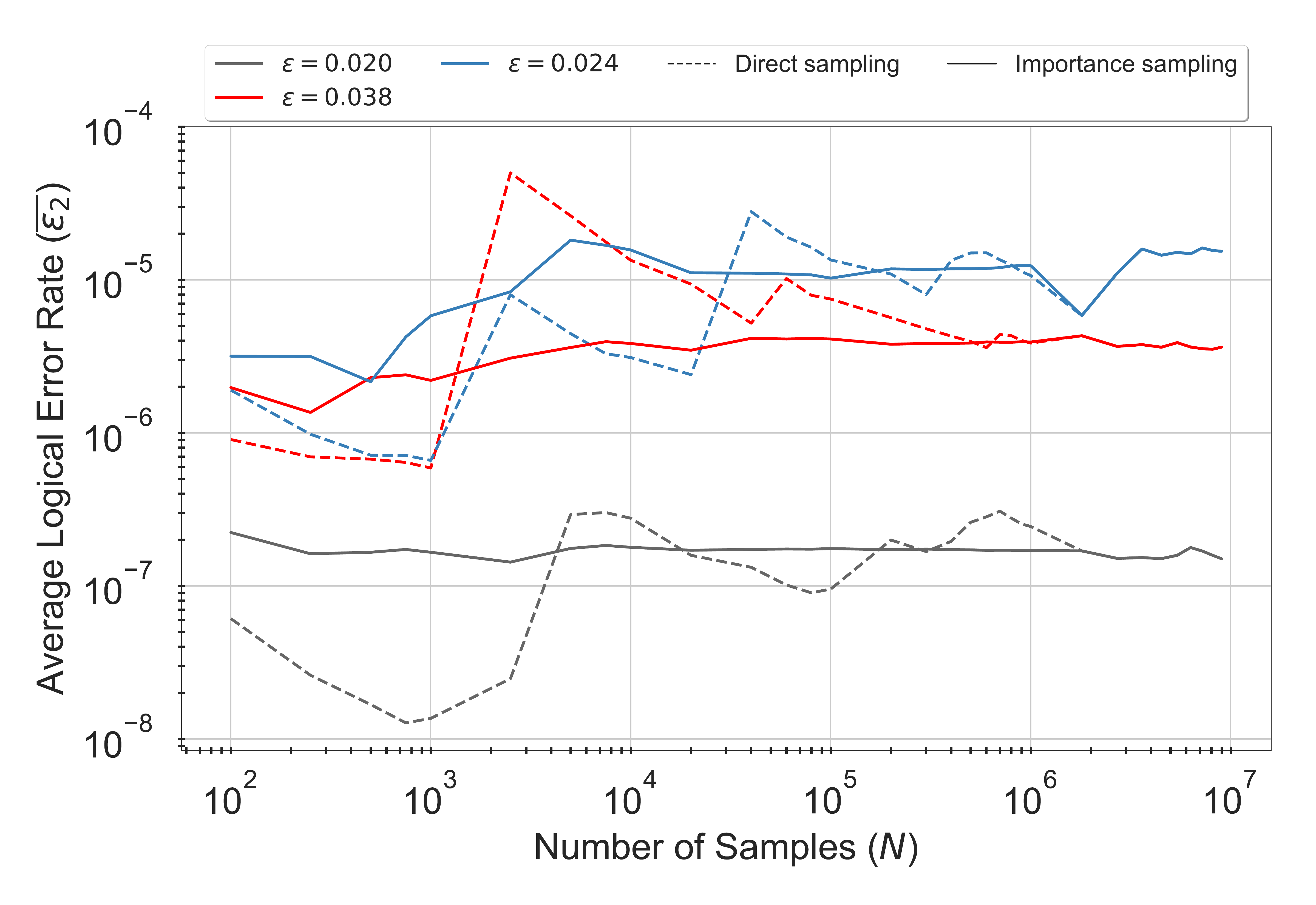}
\caption{The figure demonstrates the rapid convergence of Monte Carlo estimates of the average logical error rate for the level$-2$ concatenated Steane code using the importance sampling technique. The underlying physical noise process is the unitary error model introduced in \cref{sec:random_unitary}. The estimate from the importance sampler converges to the true average logical error rate, which is determined by direct sampling after a very large number of samples. We average the logical error rate estimates over all random unitary error ensembles, characterized by their average physical error rates. For most physical noise strengths indicated by the physical error rates, we observe that the importance sampler converges with only $1/100$ of the samples required by the direct sampler. Similar results can be found in \cite{IP17,JIBE23}.}
\label{fig:convergence}
\end{center}
\end{figure*}

We want to outline the techniques used to estimate the average logical error rate for the concatenated Steane code. Recall from our discussion in \cref{sec:qec} that the logical error rate conditioned on a syndrome measurement outcome is simply given by the infidelity of the effective channel: $r(\cE^{s}_{\ell})$, where $\ell$ denotes the concatenation level. \Cref{eq:avg_eff_chan} defines the average effective channel where the sum is over all error syndromes of the level$-\ell$ concatenated code. Recall that a level$-\ell$ concatenated code is a $[[n^{\ell}, 1, d^{\ell}]]$ code where $n$ is the total number of physical qubits in a code block and $d$ is its distance. For the level$-2$ concatenated Steane code, we find $n = 7, d = 3$ implying that there are $2^{n^{\ell}-1}$ error syndromes. Even for $\ell=2$, this is more than $10^{14}$. Hence, it is infeasible to compute the average logical error rate exactly. We instead estimate the average logical fidelity by a Monte Carlo sampling \cite{IP17,I18,IJBE22} of the syndrome distribution $\Prob(s)$ given in \cref{eq:prob_s}. We denote the sampling estimate of $r(\ol{\cE}_{\ell})$ by $r(\widehat{\cE}_{\ell})$, given by
\begin{gather}
r(\widehat{\cE}_{\ell}) = \dfrac{1}{N}\sum_{\widehat{s}}r(\cE^{\widehat{s}}_{\ell}) ~ , \label{eq:avg_infid_estimate}
\end{gather}
where $N$ is the number of syndrome samples drawn from the syndrome distribution $\Prob(s)$. 
Although it may seem straightforward to obtain $r(\widehat{\cE}_{\ell})$ in this way, there are subtleties that cause estimates to converge very slowly to the true average logical error rate due to the presence of outlier syndromes that strongly impact the average \cite{BV13,IP17,I18}. Therefore, we employ an importance sampling technique in which we sample error syndromes from a different distribution $Q(s)$ that assigns a higher weight to outlier syndromes and corrects the infidelity found for the sampled syndrome $s$ by a ratio $\Prob(s)/Q(s)$. In our simulations, we choose $Q(s) = \Prob(s)^{\alpha}$, where $\alpha$ is selected such that the uncorrectable errors, for which the average infidelity is close to one, have a weight greater than a fixed threshold \cite{JIBE23}. \Cref{fig:convergence} demonstrates how the average logical error rate estimated from the importance sampling method converges to the true average estimated from the direct sampling method using a very large number of samples. We have shown explicit convergence of the logical error rate estimates in the two sampling methods under a few unitary channels modeled by random quantum circuits in \cref{sec:random_unitary}. Clearly, the importance sampling method estimates the true average with far fewer samples. In this work, our numerical methods to estimate performance gains from different decoding strategies under various error models utilize the importance sampling method.

\end{appendix}

\bibliographystyle{plainurl}
\bibliography{refs_formatted}
\end{document}